\documentclass[12pt,twocolumn, aps,nofootinbib,preprintnumbers,superscriptaddress,numberedappendix]{openjournal}

\usepackage{natbib}

\usepackage{newtxtext,newtxmath}

\usepackage[T1]{fontenc}

\usepackage[colorlinks=true, allcolors=blue]{hyperref}

\DeclareRobustCommand{\VAN}[3]{#2}
\let\VANthebibliography\thebibliography
\def\thebibliography{\DeclareRobustCommand{\VAN}[3]{##3}\VANthebibliography}

\usepackage{graphicx}	
\usepackage{amsmath}	
\usepackage{soul}
\usepackage{microtype}
\usepackage{pythonhighlight}
\usepackage{xspace}
\usepackage{makecell}
\usepackage{pmboxdraw} 

\usepackage{lsstdesc_macros}
\usepackage{xspace}
\usepackage{graphicx}
\usepackage{hyperref}
\usepackage{url}
\usepackage{natbib}
\usepackage{tikz}
\usepackage{amssymb}
\usepackage[normalem]{ulem}
\usepackage{xcolor}
\graphicspath{{./}{./figures/}}
\usepackage{footmisc}
\usepackage{lineno}
\usepackage[normalem]{ulem}

\usepackage{tikz}
\usetikzlibrary{shapes.geometric, arrows, fit, positioning}

\tikzstyle{product} = [
    rectangle,
    minimum width=1.5cm,
    minimum height=0.7cm,
    align=center,
    draw=black,
    font=\scriptsize
]

\tikzstyle{stage} = [
    ellipse,
    minimum width=1.5cm,
    align=center,
    draw=black,
    font=\scriptsize
]

\tikzstyle{box} = [
    draw=red,
    dashed,
    thick,
    inner sep=10pt,
]

\tikzstyle{arrow} = [
    ->,
    >=stealth
]

\newcommand{\creation}{
\begin{tikzpicture}

    \node (input) [product] {Generative Model Config};
    \node (dat) [product, right of=input, xshift=3cm] {Reference Data/Priors};

    \node (train) [stage, below of=dat, xshift=-2cm, yshift=-0.5cm] {Inform Generative Model};
    \node (model) [product, below of=train, yshift=-0.5cm] {Empirical/Explicit $p(z, \mathbf{p})$};
    \node (samp) [stage, below of=model, yshift=-0.5cm] {Draw $\{z, \mathbf{p}\}$};
    \node (lat) [product, below of=samp, yshift=-0.5cm] {Underlying Galaxy Catalog\\$\{z, \mathbf{p}, p(z_t | \mathbf{p}_t)\}$};
    \node (degradation) [stage, below of=lat, yshift=-0.5cm] {Degradation};
    \node (obs) [product, below of=degradation, yshift=-0.5cm] {Biased Galaxy Sample\\$\{z, \mathbf{p}, p(z_t | \mathbf{p}_t)\}$};
    \draw [arrow] (input) -- (train);
    \draw [arrow] (dat) -- (train);
    \draw [arrow] (train) -- (model);
    \draw [arrow] (model) -- (samp);
    \draw [arrow] (samp) -- (lat);
    \draw [arrow] (lat) -- (degradation);
    \draw [arrow] (degradation) -- (obs);
    
\end{tikzpicture}
}

\newcommand{\estimation}{
\begin{tikzpicture}

    \node (input) [product] {Estimator \\ Model \\ Parameters};
    \node (prior) [product, xshift=2cm] {Prior \\ Information};

    \node (inform) [stage, below of=input, xshift=1cm] {Inform \\ Estimator};
    \node (model) [product, below of=inform, yshift=-0.5cm] {Estimation \\ Model};
    \node (data) [product, right of=model, xshift=1cm] {Test Set};
    \node (estimate) [stage, below of=model] {Apply \\ Estimator};
    \node (pdfs) [product, below of=estimate, yshift=-0.5cm] {Photo-$z$ \\ PDFs};
    \node (points) [product, right of=pdfs, xshift=1cm] {Optionally \\ requested \\ photo-$z$ \\ point \\ estimates};
    \draw [arrow] (input) -- (inform);
    \draw [arrow] (prior) -- (inform);
    \draw [arrow] (inform) -- (model);
    \draw [arrow] (model) -- (estimate);
    \draw [arrow] (data) -- (estimate);
    \draw [arrow] (estimate) -- (pdfs);
    \draw [arrow] (estimate) -- (points);
    
\end{tikzpicture}
}

\newcommand{\rail}{\code{RAIL}}

\newcommand{\ceci}{\code{ceci}}
\newcommand{\qp}{\code{qp}}
\newcommand{\tablesio}{\code{tables\_io}}
\newcommand{\cre}{\code{creation}}
\newcommand{\est}{\code{estimation}}
\newcommand{\eva}{\code{evaluation}}

\newcommand{\proj}[1]{\textsc{#1}\xspace}
\newcommand{\lsst}{\proj{LSST}}
\newcommand{\desc}{\proj{DESC}}
\newcommand{\lincc}{\proj{LINCC}}
\newcommand{\pzvc}{\proj{PZVC}}
\newcommand{\dcone}{\proj{DC1}}

\newcommand{\pz}{photo-$z$\xspace}
\newcommand{\Pz}{Photo-$z$\xspace}
\newcommand{\pzs}{photo-$z$s\xspace}
\newcommand{\Pzs}{Photo-$z$s\xspace}

\shorttitle{Redshift Assessment Infrastructure Layers (\rail)}
\shortauthors{RAIL Team et al.}

\begin{document}
\title{Redshift Assessment Infrastructure Layers (\rail): \\ Rubin-era photometric redshift stress-testing and at-scale production }

\author{The RAIL Team,}
\author{Jan Luca van den Busch$^1$}
\author{Eric Charles$^{2,3,*}$}
\author{Johann Cohen-Tanugi$^{4}$}
\author{Alice Crafford$^{5}$ }
\author{John Franklin Crenshaw$^{6,7}$}
\author{Sylvie Dagoret$^{8}$}
\author{Josue De-Santiago$^{9}$}
\author{Juan De Vicente$^{10}$}
\author{Qianjun Hang$^{11, \dag}$}
\author{Benjamin Joachimi$^{11}$}
\author{Shahab Joudaki$^{10}$}
\author{J. Bryce Kalmbach$^{2,3,7,12}$}
\author{Arun Kannawadi$^{13}$}
\author{Shuang Liang$^{2,3,14}$ }
\author{Olivia Lynn$^{5,15}$ }
\author{Alex I. Malz$^{5,15 \ddag}$}
\author{Rachel Mandelbaum$^{5,15}$}
\author{Sidney Mau$^{2,3}$}
\author{Leonel Medina-Varela$^{5}$}
\author{Grant Merz$^{16}$ }
\author{Irene Moskowitz$^{17}$}
\author{Drew Oldag$^{7,15}$}
\author{Jaime Ruiz-Zapatero$^{11}$}
\author{Mubdi Rahman$^{18}$}
\author{Markus M. Rau$^{19,20}$}
\author{Samuel J. Schmidt$^{21, \S}$} 
\author{Jennifer Scora$^{18}$}
\author{Raphael Shirley$^{22}$}
\author{Benjamin St{\"o}lzner$^{1}$}
\author{Laura Toribio San Cipriano$^{10}$}
\author{Luca Tortorelli$^{23}$ }
\author{Ziang Yan$^{1}$ }
\author{Tianqing Zhang$^{5,14,24 \P}$}
\author{the LSST Dark Energy Science Collaboration}

\thanks{Corresponding Author: $^*$ \href{mailto:echarles@slac.stanford.edu}{echarles@slac.stanford.edu}.}
\thanks{Corresponding Author: \dag~\href{mailto:e.hang@ucl.ac.uk}{e.hang@ucl.ac.uk}.}
\thanks{Corresponding Author: \ddag~\href{mailto:aimalz@nyu.edu}{aimalz@nyu.edu}.}
\thanks{Corresponding Author: \S~\href{samschmidt@ucdavis.edu}{\rm samschmidt@ucdavis.edu}.}
\thanks{Corresponding Author: \P~\href{mailto:tq.zhang@pitt.edu}{\rm tq.zhang@pitt.edu}.}

\affiliation{$^{1}$Ruhr University Bochum, Faculty of Physics and Astronomy, Astronomical Institute (AIRUB), German Centre for Cosmological Lensing, 44780 Bochum,
Germany}
\affiliation{$^{2}$SLAC National Accelerator Laboratory, 2575 Sand Hill Road, Menlo Park, CA 94025, USA}
\affiliation{$^{3}$Kavli Institute for Particle Astrophysics and Cosmology (KIPAC), Stanford University, Stanford, CA 94305, USA}
\affiliation{$^{4}$ Universit\'e Clermont-Auvergne, CNRS, LPCA, 63170 Aubière, France}
\affiliation{$^{5}$McWilliams Center for Cosmology and Astrophysics, Department of Physics, Carnegie Mellon University, Pittsburgh, PA, USA}
\affiliation{$^{6}$Department of Physics, University of Washington, Seattle, WA 98195, USA}
\affiliation{$^{7}$DIRAC Institute, University of Washington, Seattle, WA 98195, USA}
\affiliation{$^{8}$ Universit\'e Paris-Saclay, CNRS, IJCLab, 91405, Orsay, France} 
\affiliation{$^{9}$ Secihti---Departamento de F\'isica, Centro de Investigaci\'on y de Estudios Avanzados del I.P.N. Apdo. 14-740, Ciudad de M\'exico 07000, M\'exico}
\affiliation{$^{10}$ Centro de Investigaciones Energ\'{e}ticas, Medioambientales y Tecnol\'{o}gicas (CIEMAT), Av. Complutense 40, E-28040 Madrid, Spain}
\affiliation{$^{11}$Department of Physics \& Astronomy, University College London, Gower Street, London WC1E 6BT, UK}
\affiliation{$^{12}$Department of Astronomy, University of Washington, Seattle, WA 98195, USA}
\affiliation{$^{13}$Department of Physics, Duke University, Durham, NC 27708, USA}
\affiliation{$^{14}$Department of Physics and Astronomy, Stony Brook University, Stony Brook, NY 11794-3800, USA}
\affiliation{$^{15}$ LSST Interdisciplinary Network for Collaboration and Computing Frameworks, 933 N. Cherry Avenue, Tucson AZ 85721}
\affiliation{$^{16}$Department of Astronomy, University of Illinois at Urbana-Champaign, 1002 West Green Street, Urbana, IL 61801, USA}
\affiliation{$^{17}$Department of Physics and Astronomy, Rutgers, The State University of New Jersey, Piscataway, NJ 08854, USA}
\affiliation{$^{18}$ Sidrat Research, 124 Merton Street, Suite 507, Toronto, ON M4S 2Z2, Canada}
\affiliation{$^{19}$ School of Mathematics, Statistics and Physics,Newcastle University, Newcastle upon Tyne, NE17RU, United Kingdom}
\affiliation{$^{20}$ High Energy Physics Division, Argonne National Laboratory, Lemont, IL 60439, USA}
\affiliation{$^{21}$Department of Physics and Astronomy, University of California, One Shields Avenue, Davis, CA 95616, USA}
\affiliation{$^{22}$ Max-Planck-Institut f\"ur extraterrestrische Physik, Giessenbachstrasse 1, 85748 Garching, Germany}
\affiliation{$^{23}$Universit{\"a}ts-Sternwarte, Fakult{\"a}t f{\"u}r Physik, Ludwig-Maximilians-Universit{\"a}t München, Scheinerstr. 1, 81679 M{\"u}nchen, Germany}
\affiliation{$^{24}$Department of Physics and Astronomy and PITT PACC, University of Pittsburgh, Pittsburgh, PA 15260, USA}

\begin{abstract}
  Virtually all extragalactic use cases of the Vera C. Rubin Observatory's Legacy Survey of Space and Time (\textsc{LSST}) require the use of galaxy redshift information, yet the vast majority of its sample of tens of billions of galaxies will lack high-fidelity spectroscopic measurements thereof, instead relying on photometric redshifts (photo-$z$) subject to systematic imprecision and inaccuracy best encapsulated by photo-$z$ probability density functions (PDFs).
  We present the version 1 release of Redshift Assessment Infrastructure Layers (\rail), an open source \code{Python} library for at-scale probabilistic photo-$z$ estimation, initiated by the \textsc{LSST} Dark Energy Science Collaboration (\textsc{DESC}) with contributions from the \textsc{LSST} Interdisciplinary Network for Collaboration and Computing (\textsc{LINCC}) Frameworks team.
  \texttt{RAIL}'s three subpackages provide modular tools for end-to-end stress-testing, including a forward modeling suite to generate realistically complex photometry, a unified API for estimating per-galaxy and ensemble redshift PDFs by an extensible set of algorithms, and built-in metrics of both photo-$z$ PDFs and point estimates.
  \texttt{RAIL} serves as a flexible toolkit enabling the derivation and optimization of photo-$z$ data products at scale for a variety of science goals and is not specific to \textsc{LSST} data.
  We thus describe to the extragalactic science community including and beyond Rubin the design and functionality of the \texttt{RAIL} software library so that any researcher may have access to its wide array of photo-$z$ characterization and assessment tools.
\end{abstract}



\maketitle



\section{Introduction}
\label{sec:intro}

The Vera C. Rubin Observatory's Legacy Survey of Space and Time (\lsst) will obtain deep $ugrizy$ imaging of $\sim 18,000$ square degrees optimized for extragalactic science during its ten-year survey.
It will produce a catalog of 20 billion galaxies down to $i=26.4$, enabling advances in our understanding of numerous extragalactic phenomena and cosmology \citep{ivezic2019}.
Most, if not all, such studies require some notion of the distance or redshift of each galaxy, and in some cases the distribution of redshifts for ensembles of galaxies (see Table~1 of \citealt{Breivik2022}).
Redshift may be measured directly from the absorption and emission lines in a spectrum. 
However, \lsst's galaxy sample poses a twofold challenge to redshifts measured directly from the absorption and emission lines in a spectrum: 
the large number of galaxies greatly exceeds the available time on spectroscopic follow-up observing facilities, and the faint sources that comprise the bulk of the sample will be inaccessible to spectroscopic instruments regardless.

In lieu of spectroscopically confirmed redshifts, \lsst will yield photometric redshifts (\pzs) derived from broadband photometry, a standard data product of photometric galaxy surveys \citep[e.g.,][]{Tanaka2018, Bilicki2018, Buchs2019}.
\Pzs are subject to a variety of sources of inaccuracy and imprecision that are anticipated to be more severe for \lsst's faint sample, making point estimates and Gaussian uncertainties inappropriate approximations for nearly all extragalactic applications of \lsst data.
{\Pz probability density functions (PDFs) comprehensively characterize the redshift uncertainty \citep{Tanaka2018, Bilicki2018, Buchs2019} and are thus favored for LSST data.}
A thorough review of \pz uncertainty characterization can be found in \citet{newman_photometric_2022}.

The \lsst Dark Energy Science Collaboration (\lsst-\desc) aims to perform a precision cosmology analysis on LSST data and thus has stringent requirements on \pz quality \citep{the_lsst_dark_energy_science_collaboration_lsst_2018}.
A first step in the effort to achieve those goals was the experiment of \citet{schmidt_evaluation_2020}, hereafter referred to as \dcone~(Data Challenge 1), which aimed to identify the most promising \pz PDF estimators under idealized conditions to set a baseline for subsequent optimization;
machine learning codes were provided with a perfectly representative training set, and model-fitting codes were provided with the true templates used to produce the mock data set.
Its inconclusive results failed to indicate a clear winner but highlighted a few surprising discoveries, identifying nontrivial problems suggesting new priorities for \desc \pz data products:
\begin{enumerate}
    \item The dozen codes tested yielded different estimated \pz PDFs despite being provided with identical inputs, i.e., the test set and explicit prior information in the form of a training set or SED templates. 
    A reasonable explanation for the discrepancies is that the algorithms themselves impart implicit priors to the resulting \pz PDFs.
    \item The experiment evaluated multiple performance metrics used previously in the literature. 
    These metrics suggested that some real \pz estimators were outperformed by a pathological estimator that completely neglects the photometry of the test set.
    This finding shows that some metrics currently used for \pz quality assessment fail to test how well an estimator uses the information in the data to constrain redshift and thus are not appropriate performance measures.  The ability to easily compute multiple metrics appropriate for the science case at hand is an obvious requirement for any in-depth analysis.
    \item There are more principled metrics of information content of estimated \pz PDFs, but they require a notion of true \pz PDFs, rather than just true redshifts, that were not accessible with the mock data set used, nor any existing mock data set.
    Meaningful assessment of \pz PDFs can benefit from a fully probabilistic forward model with true conditional PDFs to quantify how well estimators recover this inherent uncertainty.
    \item The DC1 experiment was also a learning experience from the perspective of running a data challenge.
    To perform subsequent, at-scale tests of systematic deviations from the idealized conditions of this precursor experiment, we require robust, modular software infrastructure for creating realistically complex mock data, estimating \pz PDFs by multiple methods, and evaluating a variety of metrics, both mathematical and science case-specific.
\end{enumerate}

These discoveries led to a major revision of the DESC Science Roadmap \citep[SRM;][]{collaboration_lsst_2021} describing a new paradigm for \pz validation to enable \desc science.
The Redshift Assessment Infrastructure Layers (\rail)\footnote{\url{https://github.com/LSSTDESC/RAIL}} code was devised to address these needs for \pz data products within \desc and designed to facilitate other extragalactic \lsst science cases as well as application to other photometric data sets.
\rail is a core library used by \desc analysis pipelines, built with the support of \desc Pipeline Scientists, directable software development by \lsst's international in-kind contributors, professional software engineers from the \lsst Interdisciplinary Network for Collaboration and Computing (\lincc) Frameworks program, and astrophysics researchers at multiple career stages.
As a result of its broad applicability, it is being used as part of Rubin Observatory commissioning efforts and \pz stress-testing studies in other \lsst Science Collaborations.




This paper presents the \rail v1 release, enabling robust production and stress-testing of \pz PDFs at scale for \lsst, within and beyond \desc.
\rail was designed to address the four unmet needs enumerated  above, with the following guiding requirements:
\begin{enumerate}
    \item \rail supports fully self-consistent probabilistic modeling of redshifts and photometry, thereby providing true PDFs for comparison with estimates.
    \item \rail offers a common API to many estimators, enabling any user to conduct a comparative experiment without learning the user interface to each estimator or organizing a large group of collaborators with the required knowledge of each estimator.
    \item \rail includes principled mathematical metrics and accommodates the addition of new, science-specific metrics.
\end{enumerate}


To serve a diverse set of users with different goals, \rail provides several entry points to usage, including documentation suited to each use case.
\begin{itemize}
    \item \desc multi- and joint-probe cosmological analyses:
    \rail must form the basis of software pipelines achieving key benchmarks of cosmological constraining power for cosmic shear, large-scale structure, galaxy clusters, Type Ia supernovae, and other probes, supporting both the estimation of individual \pz PDFs and the redshift distributions of galaxy samples.
    \item Rubin-ecosystem pipeline developers: 
    \rail must enable the roadmap of the Photo-Z Validation Cooperative (\pzvc)\footnote{\url{https://dmtn-049.lsst.io/}} to produce and validate \pz estimates as part of \lsst data releases, including work to optimize decisions during commissioning. 
    \item \Pz experts with their own data and/or estimators:
    \rail is flexible enough to be used on other data.
    \item Beginners new to \pz data products:
    \rail lowers the barrier to exploring \pz estimation and validation so that even students without a local \pz expert can get started with multiple estimators, realistically complex mock data, and a variety of performance measures.
\end{itemize}

\rail is developed publicly on GitHub with continuous integration, unit tests, code reviews, and documentation.
Additionally, \rail is designed to be maximally extensible, with structural choices engineered to explicitly welcome the community contributions that are essential to its success. 

\begin{figure}
    \centering
    \includegraphics[width=1.0\columnwidth]{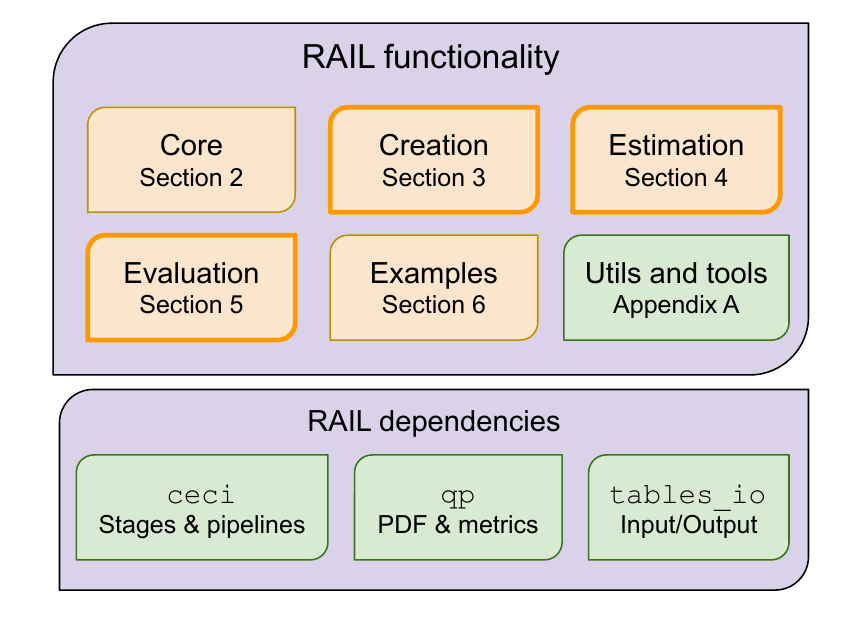}    
    \caption{Overview of structure of the \rail codebase. The core module provides building blocks for \rail with dependencies on several existing DESC software packages, i.e., \ceci, \qp, and \tablesio, as well as basic utilities and tools shared across all \rail modules. 
    The main functionality of \rail has a tripartite structure enabling experiments to optimize \pz data products, namely, creation, estimation, and evaluation {(bold blocks)}. Along with these modules, we also introduce the core functionality and examples in the main body of the paper (orange blocks). Utilities, tools, and the major dependencies of \rail (green blocks) are introduced in the Appendices.}
    \label{fig: overview}
\end{figure}

This paper is organized according to the structure of \rail codebase, as shown in Fig.~\ref{fig: overview}.
Section~\ref{sec:dep} introduces the core dependencies and outlines the libraries that comprise the \rail ecosystem.
Section~\ref{sec:cre} describes \rail's self-consistent forward-modeling functionality to create realistically complex mock photometry.
Section~\ref{sec:est} describes \rail's extensible framework for estimating \pz data products.
Section~\ref{sec:eva} describes \rail's flexible suite of metrics for evaluating \pz data products.
Section~\ref{sec:examples and tutorials} presents an end-to-end use case showcasing \rail's current functionality. We also list the existing examples for \rail. 
In each section, we present the existing functionality, outline how the community can contribute, and discuss priorities for ongoing and future development.
Section~\ref{sec:end} provides a summary, and describes the next steps for \rail development.





\section{Core Structure and Background}
\label{sec:dep}

{\rail has a variety of use cases, from creating realistic mock data (Fig.~\ref{fig:creation}), estimating  photometric redshifts (Fig.~\ref{fig:estimation flow}), to evaluating the performance of different estimation algorithms.
These functionalities are managed under a unified workflow, and their serial execution is facilitated in a traceable manner, through \rail's core structure. 
Specifically, the various functionalities are wrapped in {\it stages}, the execution in sequence is defined by {\it pipelines}, and finally, the various types of input and output data are managed by the {\it data handles}. In this section, we shall introduce in detail these building blocks of the \rail core structure.
}

{In addition to the key features presented in this Section, \rail's core structure also includes utilities and tools that facilitate easy usage of the software and simple data manipulation. These functionalities are described in Appendix~\ref{sec:util_tools}. As we shall see shortly, \rail also makes use of core dependencies developed within \desc, and their properties have guided significant aspects of \rail's design. Some of these dependencies are described in Appendix~\ref{sec: dependencies}. Finally, we briefly present the software ecosystem in Appendix~\ref{sec:eco}.
}

\subsection{Ceci Stages}
\label{sec:ceci}

The pipeline management software, \ceci\footnote{\url{https://github.com/LSSTDESC/ceci}} was developed in DESC to enable the construction of analysis pipelines from modular stages carrying provenance information for reproducibility, and to provide tools to run analyses efficiently at scale.
Much of the core functionality of \rail is built on top of \ceci.
A \ceci.\code{Stage} performs a single operation, taking a fixed set of inputs and generating a fixed set of outputs. This is done in the stage's \code{run()} function.  Stages can have configuration parameters that affect their behavior, but those parameters must be set and stored before the \code{run()} function is called to ensure reproducibility.   
The \code{run()} function itself does not take any arguments.

\rail modifies the core \ceci functionality to facilitate interactive use of stages in \textit{Jupyter} notebooks.  These are primarily implemented in \code{RailStage}, a \rail specific extension of \ceci.\code{Stage}.

The modifications are as follows:
\begin{enumerate}
\item{allowing users to interactively create stage objects by passing the configuration parameters as a \code{Python} dictionary;}
\item{implementing functions that are intended to be called interactively, such as \code{estimate()} or \code{inform()}, which, unlike \code{run()},  take arguments and  return values. 
Essentially, these functions wrap the \texttt{run()} function, and are responsible for correctly setting up the inputs and returning the correct outputs of a given stage;}
\item{implementing subclasses of \ceci.\code{Stage}, each of which has its own interactive function;}
\item{extending the functionality of the \code{DataHandle} from \code{ceci} to enable stages to correctly establish connections between themselves when they are called interactively.}
\end{enumerate}
Regarding (iv), having stages return \code{DataHandle} objects gives them a way to associate themselves with the files that they have created.
Having stages take \code{DataHandle} objects as input enables stages to know where their data are coming from, i.e., to know which stages need to be run before they can be run.

The \code{RailStage} objects defined in each of \rail's subpackages are effectively superclasses for any wrapped method;
this paper includes descriptions of these superclasses and the subclasses in the version 1 release.

\subsection{Pipelines}
\label{sec:core:pipeline}

Multiple \ceci.\code{Stage} objects may be chained together to form a \ceci.\code{Pipeline} object that is constructed as a directed acyclic graph (DAG) by connecting the inputs and outputs of the various \code{Stage} objects, thus requiring that the inputs of all stages exist or will be produced by stages that will run before them.

The core \ceci code provides mechanisms to define pipelines from \texttt{yaml} files, and to execute each stage independently, possibly under Message Passing Interface (MPI) for parallelization when running at scale.  In this {`production' mode (in contrast to `interative' mode running in, e.g., a \textit{Jupyter} notebook)}, the configuration parameters of the stages are defined in a `pipeline configuration' \texttt{yaml} file. 

To define a \rail pipeline, a user would need to have two configuration \texttt{yaml} files. The first one is a \code{ceci}  \texttt{yaml} file, which defines the global properties of the stages, such as the stage names, the classes that define each stage, their inputs/outputs, and their parallel processing parameters. The second  \texttt{yaml} file is the \rail  \texttt{yaml} file. The \rail  \texttt{yaml} file defines the parameters of each \rail stage, such as the stages they are connected to, and other stage-specific parameters. 

In the \texttt{RAIL\_pipelines}\footnote{\url{https://github.com/LSSTDESC/rail_pipelines}} package, we have made pre-built pipelines in the form of \texttt{RailPipelines} classes to generate the aforementioned configuration files. {We provide examples of \rail pipelines in Section~\ref{sec:examples and tutorials}. }


{The choice of \ceci as an effective workflow manager layer in \rail comes with some cost, the first cost being that \code{ceci} generates intermediate files in the folder where \rail is run and additional file management is needed to clean up the intermediate files. Secondly, all \rail stages are required to have output paths, which might not be users' intention when they do not want to write the output of some stages.}

\rail stages inherit their overall parallelization strategy from \ceci, employing MPI. Galaxy data can be loaded to memory in chunks, either sequentially in a single node or in parallel if several nodes are available, via \code{RailStage.input\_iterator}. This allows for processing large amounts of data while keeping memory usage to a minimum. For some algorithms that need to compute operations simultaneously over all of the data, the chunks are equally sized and distributed across all of the available nodes.

\subsection{Data Handle}
\label{sec:data_handleI}
Data in \rail are passed between stages in the form of data handles. The data handles are defined in \code{rail.core.data}. The key idea of wrapping the data in the data handle is that it allows the freedom to pass only the file name, and it allows parallel processes to partially read the file into memory. This approach drastically reduces memory usage when parallel processing large tabular data for \rail. The data handles in \rail have three main subclasses based on data types: tabular data, PDF data, and models. 

The tabular data handles (\code{TableHandle}) wrap catalog-like data, which can exist as a Numpy dictionary, pandas dataframe, pyarrow table, or Astropy table. The \code{TableHandle} interacts with \code{tables\_io} to handle different types of file formats. More information can be found about \code{tables\_io} in Appendix \ref{sec:tablesio}. The \code{qp} handle (\code{QPHandle}) wraps \code{qp} ensembles \citep{malz_qp_2017}, which are iterable data structures of 1D PDFs. Additional information about \code{qp} can be found in Appendix \ref{sec:qp}. {The model handle (\code{ModelHandle}) wraps the model generated by a particular algorithm in \rail, informed by some training data.} Regardless of their types, the models are stored in pickle files by \rail. 

The basic function of the data handle is uniform across data types. The functionalities include opening and closing a file, reading a file into memory, and writing data into files. Additionally, the tabular data handle and \code{qp} data handle can read and write in chunks for parallel processing.

\subsection{Definitions}
\label{sec: concepts}

{Before diving into the main functionalities and algorithms of \rail, we introduce key recurring concepts that recur throughout the paper for clarity. These concepts can be broadly divided into two groups: statistical and photo-$z$. We provide their mathematical notations (if applicable) and definitions in Table~\ref{tab: concepts}.  
}

\begin{table*}
\centering
\begin{tabular}{p{1.6in}p{0.4in}p{3.5in}}
\hline
 Concept & Notation & Definition\\
     \hline
     \hline
     \multicolumn{3}{l}{\hspace{-1em}{\bf Statistical Concepts}}\\
     Posterior & $p(\theta|\mathbf{d})$ & Probability of model parameters $\theta$ given the data $\mathbf{d}$. This is the output of a typical photo-$z$ estimator, where $\theta$ usually refers to redshift, and $\mathbf{d}$ is the set of photometry of the galaxy. The meaning of $\theta$ and $\mathbf{d}$ vary with photo-$z$ algorithms. In Bayesian statistics, the posterior is given by $p(\theta|\mathbf{d}) = \mathcal{L}(\mathbf{d}|\theta)\pi(\theta) / p(\mathbf{d})$, where $\mathcal{L}(\mathbf{d}|\theta)$ is the likelihood, $\pi(\theta)$ is the prior, and $p(\mathbf{d})$ is the evidence (which is of less relevance to typical photo-$z$ problems). \\
     Likelihood & $\mathcal{L}(\mathbf{d}|\theta)$ & Probability of data $\mathbf{d}$ given model parameters $\theta$. Typically used in template-fitting algorithms.\\
     Prior & $\pi(\theta)$ & Probability distribution of the model parameters, $\theta$, characterizing the prior knowledge of the inference problem. In a template-fitting algorithm, for example, this can refer to the types of SED templates used and the target redshift range. \\
     
     Photo-$z$ probability density distribution (PDF) & $p(z)$ & A photo-$z$ PDF in this paper is referred to the posterior distribution of the redshift of an {\it individual} galaxy. {This is used interchangeably with $p(z|\mathbf{p})$, where $\mathbf{p}$ is the galaxy's photometry. The conditions are omitted for notation clarity only.}\\
     Point estimate & $\hat{z}$ & A single number representing the photo-$z$ PDF, for example, the mode, median, or mean of the PDF. \\
     Ensemble redshift distribution & $n(z)$ & The normalized redshift distribution of an ensemble of galaxies. This can be constructed via the photo-$z$ PDF's or the point estimates.\\
     \hline
\multicolumn{3}{l}{\hspace{-1em}{\bf Photo-$z$ Concepts}}\\
Photometric data & $\mathbf{p}$ & Photometric data in this paper is referred to the galaxy catalog information containing photometry (i.e. fluxes or magnitudes in different filters {and colours. For some algorithms, this can also include non-photometric information, such as morphology}). This can refer to both simulated and real data.\\
Flux & $f_I$ & The amount of energy transferred in the form of photons from the source per unit area per second in a filter $I$, where $I=ugrizy$ for LSST.\\
Flux uncertainty & $\sigma_{f,I}$ & The uncertainty associated with the flux in a particular filter. In \rail, this information can either be taken from observation or provided by the degrader.\\
Magnitude & $m_I$ & The magnitude is related to flux via $m_I = -2.5\log_{10}f_I + m_0$ where $m_0$ is the zero-point magnitude off-set. For LSST, $m_0=31.4$.\\
Color & $c$& A color is defined as $c = m_I - m_{I'}$, where $I$ and $I'$ are adjacent filters. It is also often referred to as $I-I'$. For example, $u-g$ or $g-i$ color.\\
True redshift & $z_t$ & True redshift of the galaxy.\\
True redshift PDF & $p(z_t|\mathbf{p}_t)$ & The conditional PDF of the true redshift, given the true photometry of a galaxy {in the simulation, or a normalizing flow trained on a simulation}. In \rail this is provided for galaxies simulated using a probabilistic model for which the joint distribution of redshift and magnitudes from which the galaxies are sampled is known exactly. For example, this is true of galaxies simulated using \code{PZFlow}. In Section~\ref{sec:eva}, we also refer to this as $p_{\rm true}(z)$. \\
Template-fitting methods & - & Template-fitting methods typically constrain the likelihood $\mathcal{L}(\mathbf{p}|z, \Phi)$, where $\Phi$ is a set of templates. The likelihood is then turned into posterior via the prior distribution. One should hence be cautious about different priors involved in different algorithms. \\
Machine learning methods & - &  Machine learning methods typically learns the mapping between the flux space and redshifts, $p(z|\mathbf{p})$, directly via a conditional density estimator. A caveat is the non-representativeness of the training set that can bias the posterior.\\
\hline
\end{tabular}
\caption{
Key concepts and their corresponding notations and definitions mentioned in this paper.}
\label{tab: concepts}
\end{table*}





\begin{figure}
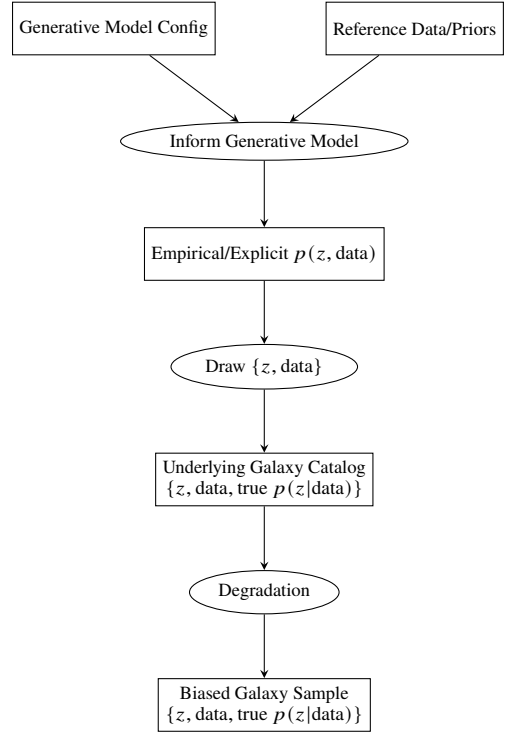

    \begin{center}
    \creation
    \end{center}
    \caption{
    {The workflow of the \rail.\cre forward modeling subpackage. Input and output data are represented by rectangles, and \rail stages are represented by ovals. 
    A typical creation pipeline starts with training a creation (`generative') model from either a reference catalog (often simulations), such as in the case of \code{PZFlow}, or from the template SEDs drawn from a prior, such as in the case of \code{DSPS}/\code{FSPS}. The empirical or explicit joint distribution of true redshifts with other catalog properties ($\mathbf{p}$) is then computed.
    A mock truth catalog (`underlying galaxy catalog') is then drawn from the creation model. This catalog is then degraded by the degradation stages, such that it mimics a noisy observed catalog (`biased galaxy sample'). }
    }
    \label{fig:creation}
\end{figure}

\section{Creation and Degradation}
\label{sec:cre}

{Mock DESC data are important for systematically testing the performance of various photo-$z$ algorithms. }
One of the lessons learned from DC1 is that it is desirable for the mock data to include not only true redshifts and \lsst photometry (i.e., fluxes in the six LSST bands) but also true posterior PDFs, $p(z_t | \mathbf{p}_t)$, which are unavailable for spectroscopically confirmed data sets as well as traditional simulations.
{This allows principled comparison with the estimated $p(z|\mathbf{p})$ under non-ideal conditions, such as  realistic noise, selection effects, and biases (e.g. see Crafford \textit{et al. in prep.}).} This is critical for the training and validation of photo-$z$ algorithms.

To address these needs, \rail.\cre enables us to create datasets with true PDFs that allow PDF-to-PDF metrics computations and forward-modeling of mock data for validating \pz approaches under realistically complex conditions. 
This is realized by two main types of stages within \rail.\cre: (1) \code{engines} that forward-model photometric catalogs and (2) \code{degraders} that modify such catalogs to introduce tunable physical imperfections.

\subsection{Engines}
\label{sec:eng}

An engine is defined by a pair of stages that are subclasses of each of the following superclasses:
\rail.\cre.\code{Modeler} makes a model of the $p(z, \mathbf{p})$ joint probability space based on input parameters or data, and \rail.\cre.\code{Creator} samples $(z, \mathbf{p})$ from the forward model. Available engines are listed  in Table~\ref{tab:eng}.


\begin{table*}
\centering
\begin{tabular}{llll}
 \hline
    \code{engine} & Approach & Home package & Reference\\
 \hline
 \hline
 \code{FSPS} & Physical & \code{rail-fsps} & \cite{Conroy2009,Conroy2010}\\
 \code{DSPS} & Physical & \code{rail-dsps} & \cite{Hearin2023}\\
 \code{pzflow} & Empirical & \code{rail-pzflow} &\cite{crenshaw2024b}\\
 \hline
\code{degrader} & Type & Home package & Reference \\
\hline\hline
\code{LSSTErrorModel} & \code{Noisifier} & \code{rail-astro-tools} & \cite{ivezic2019, crenshaw2024b}\\
\code{ObservingConditionDegrader} & \code{Noisifier} & \code{rail-astro-tools} & \cite{hang2024}\\
\code{SpectroscopicDegraders} & \code{Noisifier} & \code{rail-astro-tools} & This work\\
\code{QuantityCut} & \code{Selector} & \code{rail-base} & This work\\
\code{SpectroscopicSelectors} & \code{Selector} & \code{rail-astro-tools} & This work\\
\code{SOMSpecSelector} & \code{Selector} &
\code{rail-som} & This work\\
\code{UnrecBlModel} & \code{Degrader} & \code{rail-astro-tools} & This work\\
 \hline
\end{tabular}
\caption{
Summary of \rail.\cre.\code{engines} and degraders described in Sec.~\ref{sec:cre}.
\label{tab:eng}}
\end{table*}




\subsubsection{FSPS (Flexible Stellar Population Synthesis)}
\label{sec:fsps}

\code{FSPS} is a \rail module that creates an interface to the \code{Python} bindings of the popular stellar population synthesis (SPS) code \code{FSPS} (Flexible Stellar Population Synthesis, \citealt{Conroy2009,Conroy2010}). \code{FSPS} aims at generating realistic galaxy spectral energy distributions (SEDs) by modelling all the components that contribute to the light from a galaxy: stars, gas, dust and AGN. \code{FSPS} is widely used both for stellar population inference \citep{Johnson2021} and for forward modelling of galaxy SEDs (e.g., \citealt{Alsing2023,Tortorelli2024}).

\code{FSPS} provides substantial flexibility in terms of the prescription for modelling each of the mentioned components. It also requires physical properties of galaxies as input, such as star formation histories (SFHs), metallicities and redshift, in order to generate their SEDs. We maintained this flexibility in the interface we implemented in \code{RAIL}, allowing the user to change every possible \code{FSPS} parameter. The code has been parallelized to make efficient use of the multiprocessing capabilities of CPUs.

{The interface is integrated in the \code{RAIL} workflow, requiring as input a catalog of galaxy physical properties in the form of \code{Hdf5Handle}. These are galaxy redshifts, stellar metallicities, velocity dispersions, gas metallicities and ionization parameters (defined as the ratio of ionizing photons to the total hydrogen density), dust attenuation and emission parameters, and star-formation histories. }

\code{FSPS} follows the structure of \code{engines}. The \code{Modeler} class requires galaxy physical properties as input and produces as output an \code{Hdf5Handle} that contains the \code{FSPS}-generated rest-frame SED for each galaxy and the common rest-frame wavelength grid. The user can choose the units of the output rest-frame SEDs by setting the appropriate keyword value. {The default behavior is to output the SEDs in a wavelength grid.}

The output rest-frame SEDs constitute the input for the \code{FSPS} \code{Creator} class. The latter computes apparent $\mathrm{AB}$ magnitudes for a set of user-defined waveband filters. {Notice that the wavelength range spanned by the waveband filters should be within the SED observed-frame wavelength ranges.} A default set of filters is implemented in \code{rail.fsps}, containing the Rubin LSST filters among others.


\subsubsection{DSPS (Differentiable Stellar Population Synthesis)}
\label{sec:dsps}

\code{dsps} is a module that creates an interface in \code{RAIL} to the code \code{DSPS} (Differentiable Stellar Population Synthesis, \citealt{Hearin2023}). \code{DSPS} is implemented natively in the JAX library as its main aim is to produce differentiable predictions for the SED of a galaxy based on SPS. The implementation in JAX allows \code{DSPS} to be a factor of 5 faster than standard SPS codes, such as \code{FSPS}, and more than 300 times faster, if run on a modern GPU. \code{DSPS} does not come with stellar population templates; they must be provided by the user. The code contains a series of convenience functions that allow the user to generate stellar population templates with \code{FSPS}. If no templates are supplied, the implementation in \code{RAIL} automatically downloads a set of \code{FSPS}-generated stellar population templates.

The \code{Modeler} class of \code{dsps} requires as input a catalog of galaxy physical properties in the form of \code{Hdf5Handles}. {In particular, the user provides, for each galaxy, a star-formation history, a grid of Universe age over which the stellar mass build-up takes place, and a value for the mean and scatter of the stellar metallicity distribution. The output is an \code{Hdf5Handle} that contains galaxy rest-frame SEDs, produced over the stellar population template wavelength grid.}

The \code{Creator} class of \code{dsps} uses the output rest-frame SEDs to compute apparent and rest-frame AB magnitudes for a set of user-defined filters. Rubin-LSST filters are present in the default filter suite. The magnitudes are computed using the appropriate functions implemented in \code{DSPS} that, much like the SED generation, can take advantage of multiprocessing capabilities.

\subsubsection{\code{PZFlow} Engine} 
\label{sec:pzflow_engine}

\code{PZFlow} is a generative model that simulates galaxy catalogs using normalizing flows.
Normalizing flows learn differentiable mappings between complex data distributions and a simple latent\footnote{The simple distribution is only used for getting to the final target distribution, and is not directly accessible by the user, hence the name `latent'.} distribution, for example, a Normal distribution, hence the name \textit{normalizing} flow.
In the creation module, a normalizing flow is trained to map the distribution of galaxy colors and redshifts onto a simple latent distribution.
New galaxy catalogs can then be simulated by sampling from the latent distribution and applying the inverse flow to the samples.
In addition, because the samples are generated by sampling from a distribution we have direct access to, there is a natural notion of a \textit{true} redshift distribution for each galaxy in the catalog.
For more information, see \cite{crenshaw2024b}. Note that \code{PZFlow} is also used to perform photo-$z$ estimation, as described in Section~\ref{sec:pzflow}.



\subsection{Degraders}
\label{sec:deg}


Each \code{engine} produces a catalog from some input information, but turning the truth catalog into realistically imperfect observations necessitates additional steps in a forward model. 
A degrader may be a subclass of either \rail.\cre.\code{noisifier} (later referred to as \code{noisifier}) or \rail.\cre.\code{selector} (later referred to as \code{selector}), the first of which modifies data in place and the second of which removes rows from a catalog. 
The only exception is the blending degrader (see Sec.~\ref{sec:blending}), which changes both.
We provide several survey-specific shortcuts to mimic the selection functions of precursor data sets. Available degraders are listed  in Table~\ref{tab:eng}. 
Specifically, the \code{noisifier} superclass imposes realistically complex noise and bias to the $(z, \mathrm{photometry})$ columns, and the \code{selector} superclass introduces biased selection on the sample to mimic, e.g., an incomplete spectroscopic training sample. Fig.~\ref{fig:creation} shows the workflow of the creation sub-package.




\subsubsection{LSST Error Model}
\label{sec:modules:lsst_error_model}
The \texttt{LSSTErrorModel} is a wrapper of the \texttt{PhotErr} photometric error model \citep{crenshaw2024b}.
\texttt{PhotErr} is a generalization of the error model described in \cite{ivezic2019} that includes multiple methods for modeling photometric errors, non-detections, and extended source errors. In addition to photometric error model for LSST, we also include models for Euclid \citep{scaramella2022} and Nancy Grace Roman \citep{spergel2015} space telescopes.
The magnitude errors are estimated based on the input galaxy properties and the survey conditions, such as $5\sigma$ depth and seeing, and each galaxy has noise added to its magnitude according to a Gaussian distribution with mean zero and standard deviation equal to its magnitude error. 
For more information, see Appendix B of \cite{crenshaw2024b}.

\subsubsection{Observing Condition Degrader}
\label{sec:observing_condition}

This degrader produces observed magnitude and magnitude errors for the truth sample, based on the input survey condition maps \citep{hang2024}. The user provides a series of survey condition maps in HEALPix\footnote{\url{http://healpix.sourceforge.net}} \citep{2005ApJ...622..759G} format with specified $N_{\rm side}$, e.g. the $5\sigma$ depth in each band. The galaxies in the truth sample will be assigned survey conditions corresponding to their HEALPix pixel, either based on their coordinates in the original catalog, or randomly if only photometry is available (e.g., generated from the \code{engines}). {In the latter case, a weight map can be specified to adjust the number of galaxies assigned to each pixel.} A key input for \texttt{ObservingConditionDegrader} is \texttt{map\_dict}. This is a dictionary containing keys with the same names as parameters for \texttt{LSSTErrorModel}. Under each key, one can pass a series of paths for the survey condition maps for each band, or, if any quantity is held constant throughout the footprint, one can also pass a float number. 
The degrader then calls \texttt{PhotErr} to compute noisy magnitudes for each galaxy in each HEALPix pixel.
The output of this module is a table containing degraded magnitudes, magnitude errors, RA, Dec, and the HEALPix pixel index of each galaxy.

\subsubsection{Spectroscopic Degraders}
\label{sec:modules:spectroscopic_degraders}

\texttt{SpectroscopicDegraders} contains two simple degraders that simulate systematic errors associated with the presence of spectroscopic redshifts in spectroscopic training catalogs.

The first is \texttt{InvRedshiftIncompleteness}. It is a toy model for redshift incompleteness -- i.e., the failure of a particular spectrograph to obtain a redshift estimate for a particular set of galaxies. It takes an input catalog and {keeps all the galaxies below a configurable redshift threshold while randomly removing galaxies above it. The probability that a redshift $z$ galaxy is kept is}:
\begin{align}
    p(z) = \mathrm{min}\left( 1, \frac{z_\mathrm{th}}{z} \right),
\end{align}
where $z_\mathrm{th}$ is the threshold redshift.

The other degrader is \texttt{LineConfusion}, which simulates redshift errors due to the confusion of emission lines.
For example, if the OII line at $3727 \text{\AA}$ was misidentified as the OIII line at $5007 \text{\AA}$, the assigned spectroscopic redshift would be greater than the true redshift \citep{Newman2013}.
The inputs of this degrader are a `true' and `wrong' redshift, and an error rate.
The degrader then randomly simulates line confusion, ignoring galaxies for which the misidentification would result in a negative redshift (which can occur when the wrong wavelength is shorter than the true wavelength).

\subsubsection{QuantityCut}

This degrader provides a trimmed version of the input catalog based on selection cuts applied to the catalog quantities. The user provides the parameter \texttt{cuts}, which is a dictionary with keys being the columns to which the selection is to be applied (e.g., the $i$-band magnitude), and the values being the specific cuts. Two types of values can be provided: a single float number (e.g., 25.3), which is interpreted as a maximum value (i.e., the cut will remove samples with $i>25.3$), and a tuple (e.g., $(17, 25.3)$), which is interpreted as a range within which the sample is selected (i.e., the selected sample has $17<i<25.3$). When multiple cuts are applied at the same time, only the intersection of selected samples of each cut will be kept in the output.

\subsubsection{Spectroscopic Selectors} 

The \texttt{SpectroscopicSelection} degrader applies the selection for a spectroscopic survey. It provides tailored catalogs that match a particular spectroscopic survey for subsequent calibration steps. It can also be used to generate selected mock catalogs used as realistic reference samples. The selection criteria are cuts on magnitudes or colors adopted for the associated spectroscopic survey targeting. The current available selectors are for VVDSf02 \citep{2005A&A...439..845L}, zCOSMOS \citep{2009ApJS..184..218L}, GAMA \citep[][]{2011MNRAS.413..971D}, BOSS \citep[][]{2013AJ....145...10D}, and DEEP2 \citep[][]{2013ApJS..208....5N}.
\texttt{SpectroscopicSelection} 
requires a 2-dimensional spectroscopic redshift success rate as a function of two variables (often two of magnitude, color, or redshift), specific to the redshift survey for which selection is being emulated.
The degrader will draw the appropriate fraction of samples from the input data and return an incomplete sample.
Additional redshift cuts based on percentile can be applied when using a color-based redshift cut.

Similar functionality is provided by \texttt{GridSelection} \citep{Moskowitz2024}, which can be used to model spectroscopic success rates for the training sets used for the second data release of the Hyper Suprime Cam Subaru Strategic Program \citep[HSC; ][]{2019PASJ...71..114A}. Given a 2-dimensional grid of spectroscopic success ratio as a function of two variables (often magnitude or color), the degrader will draw the appropriate fraction of samples from the input data and return incomplete sample. Additional redshift cuts can also be applied, where all redshifts above the cutoff are removed. 
{In addition to the default HSC grid, \rail accepts user-defined setting files for the success ratio grids appropriate for other surveys.}

\subsubsection{SOMSpecSelector}

While \texttt{GridSelection} defines a selection mask in two dimensions, \texttt{SOMSpecSelector} can take any number of input features with which to define a spectroscopic selection.  This selector takes an initial complete sample (which we will call the input sample) and return a subset that approximately matches the properties of an incomplete sample (we will refer to this as the specz sample).  The selector operates by taking the list of features (which must be present in both the input and specz samples) and constructs a self-organizing map \citep[SOM; ][]{som} from the input data, creating a mapping from the higher-dimensional feature set to the 2D grid of SOM cells.  It then finds the best cell assignment for each galaxy in both the input and specz samples.  The selector builds a mask as it iterates over all cells, and for each cell returns a random subset of input objects that lie in that cell that equal in number to specz objects in the cell.  If the cell has more specz objects than are available in the input catalog, then it returns all that are available.  By matching the number of objects cell by cell the selector naturally mimics the features of the specz sample.

\subsubsection{Blending Degrader}
\label{sec:blending}

This degrader creates mock unrecognized blends based on source density. Unrecognized blends are sources overlapping too closely in projection and are detected as one object (referred to as `ambiguous blends' in \citealt{dawson}). This degrader first searches for close objects that are likely to become unrecognized blends, then merges their fluxes to create one blended object. The source IDs of blend components are saved for references.

{The blending components} are found by matching the RA and Dec coordinates of an input catalog with itself via a Friends-of-Friends (FoF) algorithm \citep{FoF}. The advantage of the FoF algorithm is that it can produce unrecognized blends from multiple sources rather than just pairs. The algorithm groups sources such that within each group, every source is separated from at lease one another group member by an angular distance less than a specified `linking length'. {By setting a small enough linking length (e.g., 1 arcsec), we assume that all group members will be blended into one detection. In the future, we might implement options for a more sophisticated identification of blends using source sizes and shapes.} In the current release, this degrader simply sums up fluxes over all group members to create one blended object per group. {Note that we do not currently simulate the impact on aperture photometry due to irregular profiles of blends either, but are motivated to conduct such a study in the future.}

Note that the truth redshifts of blended objects are ambiguous since they are composed of multiple objects. We provide several summary columns for the truth: \texttt{z\_brightest} is the redshift of the brightest component; \texttt{z\_mean} is the average redshift of all components; and \texttt{z\_weighted} is the flux-weighted average redshift. For blended objects composed of more than (including) two components, the standard deviation of redshifts is provided. The decision on the truth redshift is left to the users. For more complicated truth estimation -- e.g., considering the colors of components, as bluer galaxies tend to have strong emission lines which are often used to infer redshifts from spectroscopy -- users have the option to trace the components with source IDs. The tutorial \textbf{blending\_degrader\_demo} illustrates how to match the output catalog with the source IDs and the input catalog to access more information.

The order of application is particularly important for this degrader. Generally, this degrader should be applied before any selections on the truth catalog, including any magnitude, color, or signal-to-noise ratio cuts. The reason is that bright sources can blend with fainter ones, and two faint sources might blend into a brighter object that enters the target depth selection. For example, a magnitude difference of $\sim2.5$ translates roughly into a flux contamination of 10\%. However, applying this degrader to the original truth catalog without any cuts can be a computational burden, because the truth catalog is often much larger than the target-depth catalog. To mitigate this issue, one can use a magnitude cut to decrease the target depth by {an arbitrary threshold (e.g., 2 or 3 magnitudes)} before running this degrader. 

While preliminary studies have addressed some aspects of blending on \pz~\citep[e.g.,][]{2022MNRAS.514.5905N}, a thorough quantitative exploration of this topic will be important to develop a deeper understanding of the issue and its impacts on various science cases.


\begin{figure}
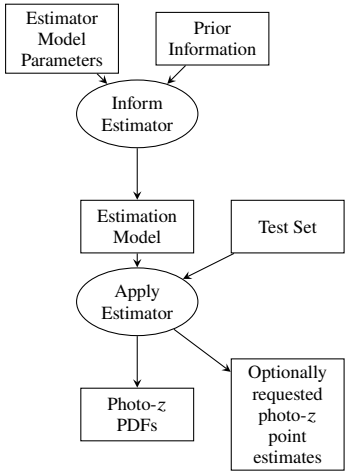

    \begin{center}
    \estimation
    \end{center}
    \caption{
    The workflow of a typical estimation \rail pipeline. The training data and prior information are fed into the Informer, which generates the photo-$z$ model. Then the model is combined with the test dataset to produce the photo-$z$ PDFs. Optional point estimate of the PDF can be requested during the estimation. Similarly, to Fig.~\ref{fig:creation}, input and output data are represented by rectangles, and \rail stages are represented by ovals.
    }
    \label{fig:estimation flow}
\end{figure}

\section{Photo-$z$ Estimation}
\label{sec:est}

\rail.\est encompasses all methods that derive redshift information from photometry, as either an estimate of per-galaxy \pz PDFs, a summary of the redshift distribution $n(z)$ for an ensemble of galaxies, or tomographic bin assignments. Technically, information other than photometry can also be input to the photo-$z$ algorithms and is allowed in \rail, especially for the machine learning methods.
Every such method is implemented with an \code{Informer} stage paired with any combination of \code{Estimator}, \code{Summarizer}, and \code{Classifier}, depending on which procedures are supported by the underlying estimator and wrapped for \rail.
An \code{Estimator} produces a \qp.\code{Ensemble} of per-galaxy \pz PDFs, a \code{Summarizer} produces a \qp.\code{Ensemble} of redshift distributions and/or samples thereof, and a \code{Classifier} produces per-galaxy integer class IDs {for tomographic binning}. An \code{Informer} generates a model for the \code{Estimator}, \code{Summarizer}, and \code{Classifier} by the training data.
Because \ceci requires stages to have fixed numbers and types of inputs, each of these stage types is implemented in at least one flavor specifying what it takes as input; 
\code{CatInformer} and \code{CatEstimator} take as input a photometric galaxy catalog with magnitudes; \code{PZInformer}, \code{PZClassifier}, and \code{PZSummarizer} take as input a \qp.\code{Ensemble} of per-galaxy \pz PDFs; and \code{SZPZSummarizer} takes as input both a spectroscopic galaxy catalog and a \qp.\code{Ensemble} of per-galaxy \pz PDFs. Specific algorithms, which are detailed below, are implemented as subclasses of these parent classes.

{Fig.~\ref{fig:estimation flow} shows the flow chart of estimation algorithms.}

\begin{table*}
\centering
\begin{tabular}{llll}
 \hline
    Algorithm name & Available stages & Home package & Reference\\
 \hline
 \hline
 \code{BPZ} & \code{CatInformer}, \code{CatEstimator} & \href{https://github.com/LSSTDESC/rail_bpz}{\code{rail-bpz}} & \citet{Benitez:2000}\\
 \code{CMNN} & \code{CatInformer}, \code{CatEstimator} & \href{https://github.com/LSSTDESC/rail_cmnn}{\code{rail-cmnn}} & \citet{Graham:2018}\\
 \code{Delight} & \code{CatInformer}, \code{CatEstimator} & \href{https://github.com/LSSTDESC/rail_delight}{\code{rail-delight}} & \citet{Leistedt:2017}\\
 \code{DNF} & \code{CatInformer}, \code{CatEstimator} & \href{https://github.com/LSSTDESC/rail_dnf}{\code{rail-dnf}} & \citet{2016MNRAS.459.3078D}\\
 \code{FlexZBoost} & \code{CatInformer}, \code{CatEstimator} & \href{https://github.com/LSSTDESC/rail_flexzboost}{\code{rail-flexzboost}} & \citet{Izbicki:2017}\\
 \code{GPz} & \code{CatInformer}, \code{CatEstimator} & \href{https://github.com/LSSTDESC/rail_gpz_v1}{\code{rail-gpz-v1}} & \citet{Almosallam:2016}\\
 $k$-nearest neighbors & \code{CatInformer}, \code{CatEstimator} & \href{https://github.com/LSSTDESC/rail_sklearn}{\code{rail-sklearn}} & This work\\
 \code{LePHARE} & \code{CatInformer}, \code{CatEstimator} & \href{https://github.com/LSSTDESC/rail_lephare}{\code{rail-lephare}} & \citet{1999MNRAS.310..540A}\\
 Neural network & \code{CatInformer}, \code{CatEstimator} & \href{https://github.com/LSSTDESC/rail_sklearn}{\code{rail-sklearn}} & This work\\
 \code{pzflow} & \code{CatInformer}, \code{CatEstimator} & \href{https://github.com/LSSTDESC/rail_pzflow}{\code{rail-pzflow}} & \cite{crenshaw2024b}\\
 Random Gaussian & \code{CatInformer}, \code{CatEstimator} & \href{https://github.com/LSSTDESC/rail_base}{\code{rail-base}} & This work\\
 \code{TPZ} & \code{CatInformer}, \code{CatEstimator} & \href{https://github.com/LSSTDESC/rail_tpz}{\code{rail-tpz}} & \citet{Carrasco-Kind:2013}\\
 \code{trainZ} & \code{CatInformer}, \code{CatEstimator} & \href{https://github.com/LSSTDESC/rail_base}{\code{rail-base}} & \citet{schmidt_evaluation_2020}\\\hline
 Uniform binning & \code{PZClassifier} & \href{https://github .com/LSSTDESC/rail_base}{\code{rail-base}} & This work\\
 Equal count binning & \code{PZClassifier} & \href{https://github.com/LSSTDESC/rail_base}{\code{rail-base}} & This work\\
  Random forest & \code{CatInformer}, \code{CatClassifier} & \href{https://github.com/LSSTDESC/rail_sklearn}{\code{rail-sklearn}} & \cite{2001MachL..45....5B}\\\hline
 Variational inference stacking & \code{PzInformer}, \code{PZSummarizer} & \href{https://github.com/LSSTDESC/rail_base}{\code{rail-base}} & \cite{Rau:2022}\\
  \code{minisom} & \code{CatInformer}, \code{PZSummarizer} & \href{https://github.com/LSSTDESC/rail_som}{\code{rail-som}}& This work\\
 Naive stacking & \code{PzInformer}, \code{PZSummarizer} & \href{https://github.com/LSSTDESC/rail_base}{\code{rail-base}} & \cite{2020arXiv200712178M}\\
\code{somoclu}& \code{CatInformer}, \code{PZSummarizer} & \href{https://github.com/LSSTDESC/rail_som}{\code{rail-som}}& This work\\
\code{NZDIR} & \code{CatInformer}, \code{CatSummarizer} & \href{https://github.com/LSSTDESC/rail_sklearn}{\code{rail-sklearn}} & \cite{Lima:2008}\\
 Point estimate histogram & \code{PzInformer}, \code{PZSummarizer} & \href{https://github.com/LSSTDESC/rail_base}{\code{rail-base}} & This work\\
 \code{yet\_another\_wizz} & \code{YawSummarize} (final stage) & \href{https://github.com/LSSTDESC/rail_yaw}{\code{rail-yaw}} & \citet{vandenbusch20}\\
 \hline
\end{tabular}
\caption{
Summary of the pre-wrapped estimators/summarizers/classifiers described in Sec.~\ref{sec:est}.}
\label{tab:alg}
\end{table*}
\subsection{Machine Learning-based Catalog Estimators}
\label{sec:ML-estimators}

\subsubsection{CMNN (Color-Matched Nearest Neighbor)}
\label{sec:cmnn}
\code{CMNN}, short for {\textit{Color-Matched Nearest Neighbor}}, is a method introduced in \citet{Graham:2018}.  The algorithm identifies nearest neighbors based on the Mahalanobis distance in color space from a set of galaxies with known spectroscopic redshifts, where the Mahalanobis distance, $D_M$, between the test galaxy and a single training galaxy is defined as:
\begin{equation}
    D_M = \sqrt{\sum^{N_{\rm colors}} \frac{(c_{\rm train}-c_{\rm test})^2}{(\delta c_{\rm test})^2}},
\end{equation}
where $N_{\rm colors}$ is the number of colors available, $c_{\rm train}$ are the colors of a single training galaxy, $c_{\rm test}$ are the colors of the test galaxy, and $\delta c_{\rm test}$ are the color uncertainties for the test galaxy, computed for each $c_{\rm test}$ color by adding in quadrature the magnitude uncertainties of the two magnitudes used to define the color. Neighboring galaxies within a minimum Mahalanobis distance, defined via the percent point function (PPF), are retained, and there are several options from which a user can estimate a PDF from this subset: 1) a single galaxy from the subset is chosen at random from the subset; 2) a single galaxy is chosen, but with a probability weighted by the inverse of the square root of Mahalanobis distance; 3) the galaxy withthe smallest Mahalanobis distance is chosen.  In all three instances, the PDF for a galaxy is returned as a single Gaussian, where the central value is assigned to the spectroscopic redshift of the galaxy chosen from one of the three options listed above, and the uncertainty is calculated by computing the standard deviation of all galaxies in the minimum distance subset. When there are less than $n_{\rm min}$ galaxies in the subset, the redshift will fail and an error flag is assigned to the galaxy. 

\subsubsection{DNF (Directional Neighborhood Fitting)}\label{sec:dnf}

{\code{DNF} (Directional Neighborhood Fitting) is a photometric redshift estimation method described by \citet{2016MNRAS.459.3078D}. The algorithm estimates the \pz of each galaxy from the hyperplane that best fits its directional Neighborhood in the training sample. \code{DNF} supports three main distance metrics: \code{ENF} (Euclidean Neighborhood Fitting), \code{ANF} (Angular Neighborhood Fitting), and a combination of both (\code{DNF}). \code{ENF} relies on the Euclidean distance, making it a straightforward and commonly used approach in k-Nearest Neighbors (\code{kNN}) methods. \code{ANF} uses a normalized inner product, which provides the most accurate redshift predictions, particularly in data sets with fluxes in more than four bands and sufficiently high signal-to-noise ratios.  Finally, \code{DNF} combines the Euclidean and angular metrics, improving accuracy in cases of few bands and low signal-to-noise conditions.}

{\code{DNF} provides two photometric redshift estimates: \code{DNF\_Z}, which is computed as the weighted average or hyperplane fit of a set of neighbors determined by a specific metric, and \code{DNF\_ZN}, which corresponds to the redshift of the closest neighbor and can be used for estimating the sample redshift distribution.}

{To construct the PDF for photometric redshifts, \code{DNF} selects a set of nearest neighbors based on one of these distance metrics and assigns weights to them. The PDF is computed by estimating the redshift distribution of the selected neighbors and applying a Gaussian smoothing function to account for uncertainties.}

\subsubsection{FlexZBoost}
\label{sec:estimation:flexzboost}

\code{FlexZBoost} \citep{Izbicki:2017, Dalmasso:2020} is an algorithm based on conditional density estimate that uses the \code{FlexCode}\footnote{available at \url{https://github.com/lee-group-cmu/FlexCode}} package.  The package parameterises the PDF as a linear combination of orthonormal basis functions (a set of unit vectors in the color space that are orthogonal to each other), where the basis function coefficients can be determined by regression.  The \rail implementation uses  \code{xgboost}\citep{Chen:2016} to perform the regression.  The basis function representation of the photo-z PDF of a galaxy can lead to small-scale residual `bumps'. In the course of training the density estimate, an optimal threshold (configuration parameter \code{bump\_thresh}) below which small-scale features are removed is determined by setting aside a fraction of the training data and minimizing the CDE loss at different threshold values.  Additionally, the width of the final PDF is similarly optimized by the inclusion of a `sharpening' parameter that scales the PDF by a power law value $\alpha$.  Again a fraction of the training data is set aside and the CDE loss is minimized over a set of $\alpha$ values.  The resultant \pz PDF distributions can be stored as \code{qp.Ensembles} either in their native basis function representation, or as a linearly interpolated grid.

\subsubsection{GPz}
\label{sec:gpz}
\code{GPz} is an algorithm based on sparse Gaussian Process, which was introduced by \citet{Almosallam:2016}. The current \rail implementation of \texttt{GPz} is the preliminary version; that is, it predicts a single Gaussian PDF rather than the more sophisticated multimodal PDFs that are implemented in newer versions of \texttt{GPz} \citep{Stylianou2022}.  \texttt{GPz} models both the mean and standard deviation of the Gaussian PDF as a linear combination of basis functions, and learns the parameters for the basis functions via a Gaussian process.  The method can make one of several assumptions about the covariance between these basis functions, which are controlled via the configuration parameter \code{gpz\_method} as outlined in the \rail documentation.

\subsubsection{\code{PZFlow} Estimator}
\label{sec:pzflow}
\code{PZFlow} uses normalizing flows for photo-$z$ estimation.
It takes a catalog of galaxy colors and redshifts and learns a differentiable mapping from the data space to a simple latent space, such as a Normal distribution.
Once trained, the flow can then be used to estimate the likelihood of any redshift and color combination using the equation:
\begin{align}
    p_\mathrm{data}(z, \mathrm{colors}) = p_\mathrm{latent}(f(z, \mathrm{colors})) \,  |\mathrm{det}\nabla f(z, \mathrm{colors})|,
\end{align}
where $f$ is the action of the flow.
A photo-$z$ posterior can then be estimated by evaluating this probability over a grid of redshifts, and normalizing the posterior to unit probability.
See \cite{crenshaw2024b} for more details.

\subsubsection{Scikit-Learn methods}
\label{sec:modules:knn}
Two of the estimator codes that depend on \texttt{scikit-learn} \citep{scikit-learn} (included in the \code{rail\_sklearn} package) : a nearest-neighbor estimator and a neural network estimator.  The nearest-neighbor code estimates redshift PDFs as a Gaussian mixture model, where the number of Gaussians, $M$, is determined during the inform stage, as are the width of the Gaussians. This is done by setting aside a fraction of the training data as a validation set and minimizing the Conditional Density Estimate (CDE) Loss of the PDFs versus the true values for that set.  \code{KNearNeighInformer} uses \code{sklearn.neighbors.KDTree} to build a tree from the colors, or colors plus a reference band magnitude, of the training data.  \code{KNearNeighEstimator}  then searches the tree for the $M$ closest neighbors, and constructs a PDF with $M$ Gaussians centered at each of the corresponding nearest neighbor redshifts.

The neural network estimator is an unsophisticated implementation and is not meant to be a competitive algorithm. Instead, it is used as a simple example code and a baseline against which to test. This method constructs a model using \code{sklearn.neural\_network.MLPRegressor} to build a neural network trained on one magnitude (set by the \code{ref\_band} configuration parameter) and all of the colors from the training data, though it first regularizes the data using \code{sklearn.preprocessing.StandardScaler.transform}.  The network is set up using two hidden layers of size twelve, and a hyperbolic tangent activation function.   The estimation stage produces a Gaussian redshift PDF by running the \code{MLPRegressor}'s \code{predict} method to estimate the mean redshift.  A configuration parameter, \code{width} is used to set the width of the Gaussian PDF, which is scaled by $(1+z)$ to increase with redshift, since the uncertainty in wavelength, which directly translate to photo-$z$ uncertainty, scales with $(1+z)$.

\subsection{Template-based Catalog Estimators}
\label{sec:template-estimators}

\subsubsection{BPZ (Bayesian Photometric Redshifts)}
\label{sec:estimation:bpz}

\code{BPZ} is a template-based estimator developed by \cite{Benitez:2000}, see that reference and \cite{Coe:2006} for more details on the algorithm.  Like many template-based codes, it operates by computing synthetic fluxes for an input set of SEDs by integrating the products of the SEDs and the filter bandpass curves for a particular survey.  For each SED the algorithm computes a $\chi^2$ value using the (scaled) fluxes and the observed data.  These $\chi^2$ values are converted into likelihoods, and an SED-type-dependent apparent magnitude prior is applied to the likelihoods before they are marginalized over type to produce a final posterior probability distribution. Two additional quantities are included in the ancillary data: \code{tb} corresponding to the `best-fit SED type' (evaluated at the mode redshift), and \code{todds}, a parameter that gives the fraction of the probability that comes from SED type \code{tb} at the mode redshift.  Low values of \code{todds} may be indicative of a poor fit.




\subsubsection{LePHARE}
\label{sec:estimation:lephare}

We have also implemented the LePHARE code within \rail. The Photometric Analysis for Redshift Estimation code \citep[LePHARE:][]{1999MNRAS.310..540A,2006A&A...457..841I} is a template-fitting algorithm written in C++ with a \code{Python} wrapper that can be used to estimate redshift and  physical property posterior. We have implemented it within \rail with a default set of parameters optimised for LSST passbands, but it is fully customisable as per LePHARE in general with configuration parameters that are extensive and well documented. These default configurations are based on those used for the COSMOS2020 data sets \citep{Weaver_2022}. The full set of values are available in the public version of the LePHARE code. This adds functionality such as the estimation of stellar mass, star-formation rate, and best-fitting model.

\subsection{Hybrid Catalog Estimators}
\label{sec:hybrid-estimators}

\subsubsection{Delight}
\label{sec:delight}
\cite{Leistedt:2017} introduced a novel approach to inferring photometric redshifts which combines some of the strengths of machine learning and template-fitting methods by implicitly constructing flexible template SEDs directly from the spectroscopic training data, called Delight.
It is a method for calculating the posterior probability of redshift given a catalog of deep observations acting as a data-driven prior. 
The catalog can have observations in arbitrary bands and with arbitrary noise; Gaussian processes are used as a principled method to implicitly construct  SEDs (capturing the effects of redshifts, bandpasses and noise).
The hyperparameters of the Gaussian process can be optimized as a calibration step.


\subsection{Image-based Estimators}
\label{sec:imae-estimators}

\subsubsection{DeepDISC}
\label{sec:deepdisc}
Detection, Instance Segmentation and Classification with deep learning \citep[\code{DeepDISC};][]{Merz23,Merz2024} is a framework that utilizes instance segmentation models developed for computer vision research.  \code{DeepDISC} is built on \code{detectron2}, a toolkit and repository for instance segmentation models.  \code{DeepDISC} models are composed of a backbone network that extracts features from the input images, a Region Proposal Network (RPN) that localizes objects in the images and Region of Interest (ROI) Heads that perform per-object measurements.  The \rail implementation of \code{DeepDISC} contains a network in the ROI Heads that parametrizes the redshift PDF as a Gaussian mixture model by using a Mixture Density Network (MDN).  The number of Gaussian components of the MDN can be set by the user.  Object redshifts are learned by minimizing the negative log likelihood loss function given the training sample redshifts and MDN model.

The \code{CatInformer} stage of \code{DeepDISC} requires that the images and corresponding metadata are stored in HDF5 format.  Utility functions within the \code{data\_format.conversions} module of \code{DeepDISC} will create the HDF5 files from \texttt{json} metadata (also produced by \code{DeepDISC}) and images stored in \texttt{fits} format or as \texttt{numpy ndarrays}.  Each row of the files contains a corresponding flattened image or dictionary of metadata.  Metadata includes per-image information including image shape and world coordinate system, and per-object information including bounding box coordinates, redshift, and an optional segmentation mask.  The \code{CatEstimator} stage of \code{DeepDISC} will output a \code{qp.Ensemble} that contains redshift PDFs along with ancillary information including object RA and Dec, and a detection confidence score. {A set of example images with redshifts and metadata is available in the \code{DeepDISC} repository\footnote{\url{https://github.com/burke86/deepdisc}}.  Images were produced using the BlendingToolkit \citep{Mendoza25}, a simulation framework designed to test galaxy detection and deblending.}


\subsection{Summarizers}
\label{sec:summarizers}


{Here we describe methods that can summarize the redshift distribution of an ensemble, whether based on \pz or on other dataset such as spectroscopic redshift, or both. The calibration modules, which make adjustments globally to photo-$z$ based on extra information from other datasets, usually reference samples of a spectroscopic survey, also are also among the summarizers. }

\subsubsection{NZDir}
\label{sec:nzdir}
The \code{NZDir} algorithm is an implementation of the `direct' calibration method described in \citet{Lima:2008} and used in the KiDS-450 analysis \citep{2017MNRAS.465.1454H, 2020A&A...633A..69H}.  The algorithm is a direct calibration in that it attempts to find the closest training galaxy to each photometric galaxy in parameter space, and increments the redshift histogram at the training redshift for each photometric galaxy to build up the ensemble redshift estimate representing the entire sample.  For computational efficiency, \code{NZDir} actually does this in reverse, iterating over each spectroscopic galaxy and finding nearby photometric galaxies.  In more detail, for the inform stage, \code{NZDirInformer} uses \code{sklearn.neighbors.NearestNeighbors} to determine  the Euclidean distances to the $k$-th nearest neighbor (set by configuration parameter \code{n\_neigh}) from amongst the entire training set, and stores this in the model.  The summarizer stage, \code{NZDirSummarizer}, then builds a KDTree of all of the photometric data, and for each spectroscopic galaxy, finds all photometric galaxies within the distance to the $k$-th neighbor calculated in the inform stage, and increments a redshift histogram in the bin containing the training redshift.  The algorithm has optional weights for both the training and photometric samples that can scale the increment value in order to account for incompleteness or other systematic effects.  It should be noted that, while this process of iterating over the spectroscopic sample rather than the photometric sample is more efficient, it can miss some photometric galaxies that do not fall within the $k$-th nearest neighbor distance of any of the training set objects, and thus could introduce bias in the final redshift estimate. This deficiency is addressed in the \code{minisom} and \code{somoclu} summarizers described below.  The final output is an ensemble of redshift estimates consisting of a histogram parameterization, one for each of the $N$ bootstrap samples constructed.


\subsubsection{Self-Organizing Map (SOM)}
\label{sec:som}
The Self-Organizing Map (SOM) is an unsupervised machine learning algorithm that maps high-dimensional vectors to cells on a two-dimensional map while preserving the topological properties of the high-dimensional vectors by faithfully maintaining the distance between these data vectors. The basic idea of SOM-based redshift calibration is to construct a SOM with galaxy colors or magnitudes and assume that galaxies mapped into the same cell share the same properties such as $n(z)$ and galaxy types. Therefore, we can use a spectroscopic galaxy sample as a reference and map it onto the same SOM, and use the $n(z)$ of the reference galaxies to represent that of the photometric galaxies in the same cell. The $n(z)$ of the whole photometric sample is estimated by combining the $n(z)$ distributions of all the cells.

\code{rail\_som} contains two implementations of SOM-based calibration: \code{minisom\_som} based on a light minimalistic SOM package \code{minisom}\footnote{\href{https://pypi.org/project/MiniSom/}{https://pypi.org/project/MiniSom/}}, and \texttt{somoclu\_som} with the \texttt{somoclu} package\footnote{\href{https://somoclu.readthedocs.io/en/stable/}{https://somoclu.readthedocs.io/en/stable/}}. \texttt{somoclu} is a parallelized package that can construct SOM on large datasets. It supports rectangular and hexagonal SOM cells, a planar and toroidal topologies, and a random or principal component analysis initialization. There is an option to further group the SOM cells into hierarchical clusters using the \texttt{AgglomerativeClustering} class from the \texttt{sklearn.cluster} package. { This option adds the flexibility and speed to group galaxies in the magnitude/color space. } 

The calibration process follows \citet{Wright_2020} in general. In the inform stage, the SOM is trained on magnitudes or colors, or a mixture of them, and then the SOM cells are grouped into clusters. The informer will attach the cluster index of each input source in a new column. The summarizer will then summarize the redshift distribution based on the redshift distribution of reference galaxies in each cell or cluster. It can also calculate the effective number density of weighted photometric sources as defined in \citet{2012MNRAS.427..146H}, which can be used to evaluate how well the photometric sample is represented by the reference spectroscopic sample. The weights often depends on the science cases, e.g., for weak lensing, the weights are function of the signal-to-noise and galaxy shapes \citep{Mandelbaum2018}. 

To select sources that are well represented by the reference sample, we also incorporate the quality cut (QC) defined by \citet{Wright_2020} to rule out SOM cells or clusters in which the target sources are badly represented by the reference sample. The current criteria implemented are based on QC1: the difference between photometric redshift and the true redshift of the reference sample needs to be small enough to rule out catastrophic outliers; QC2: the difference between the mean photometric redshift of the target sample and the reference sample is small enough to rule out regions in the color space where the target sample is not well represented.


\subsubsection{Naive methods}
\label{sec:naive_methods}

The \texttt{NaiveStackSummarizer} takes the photo-$z$ PDF from each catalog object and simply average them over the ensemble to produce the ensemble redshift distribution $n(z)$, i.e., $n(z) = \sum_i^N p_i(z)/N$. A set of bootstrap realizations is also produced to estimate the uncertainties on the distributions. The \texttt{NaiveStackMaskedSummarizer} can deal with tomographic bins. Given the tomographic bin file, the stage loops through galaxies in each tomographic bin specified by a bin mask and produces the $n(z)$ for each tomographic bin. As pointed out in Appendix A of \citet{Rau2023} this ``stacking" is a valid (though not necessarily optimal) ensemble estimate for CDE-based methods, e.g., \code{FlexZBoost}.  However, this is a `naive' way to obtain the ensemble redshift distribution; as pointed out in \cite{2020arXiv200712178M}, it overestimates the photometric redshift uncertainty.

The naive methods can serve as baselines when making comparisons with more sophisticated methods. They also serve as good algorithms for unit testing. 

\subsubsection{Yet Another Wizz (YAW)}
\label{sec:yet_another_wizz}

All the calibration methods mentioned above rely on photometric properties of galaxies or properties of the population color space. Cross-correlation or clustering redshifts (CCs) represent an independent approach to redshift calibration by leveraging the spatial clustering of galaxies, measured from the amplitude of the angular cross-correlation function, $w_{\rm ru}(z)$, between a (typically spectroscopic) reference sample and a sample with unknown redshift distribution  \citep{Newman08,Schmidt13,Menard13,Gatti18,vandenbusch20}. The relationship between the unknown redshift distribution $n_{\rm u}(z)$ and the cross-correlation function can be expressed by
\begin{equation}
  \label{eq:crosscorr}
  n_{\rm u}(z) \propto \frac{w_{\rm ru}(z)}{\sqrt{w_{\rm rr}(z) \, w_{\rm uu}(z)}} \, ,
\end{equation}
where $w_{\rm rr}(z)$ and $w_{\rm uu}(z)$ (angular autocorrelation function amplitudes of the reference and unknown sample) parameterise the redshift evolution of the galaxy bias.

The method proposed in \citet{Schmidt13} (measuring the correlation functions between pairs of photometric samples and reference samples in a single bin of radial distance between the two samples of fixed physical scale) is implemented in \code{yet\_another\_wizz}\footnote{\url{https://github.com/jlvdb/yet_another_wizz}} (YAW, \citealt{vandenbusch20}), for which we provide a wrapper in \code{cc\_yaw}. This wrapper consists of a number of stages that interface all primary YAW functionality:
\begin{itemize}
    \item data preparation (\code{YawCacheCreate}), i.e., splitting input data samples into regions for spatial resampling and covariance estimation,
    \item measurement of the angular autocorrelation function amplitude (\code{YawAutoCorrelate}) to estimate the evolution of galaxy bias with redshift,
    \item measurement of the angular cross-correlation amplitude (\code{YawCrossCorrelate}), and
    \item estimation of the ensemble redshift distribution (\code{YawSummarize}) according to Eq.~\eqref{eq:crosscorr}.
\end{itemize}

Two challenges for clustering redshifts are the estimation of the galaxy bias evolution, i.e., a dependency of how galaxies trace the large-scale strucutre on redshift, of the unknown sample ($w_{\rm uu}$) and the fact that clustering redshift estimates, by definition, are not a probability density but a ratio of correlation functions (see Eq.~\ref{eq:crosscorr}). YAW does not address these issues directly and therefore its output needs to be modeled accordingly. Nevertheless, clustering redshifts provide a powerful, independent estimate on the redshift distribution $n(z)$ that may be used additionally as prior information or in combination with photometric redshift estimation methods.

\subsection{Classifiers}
\label{sec:classifiers}


The Classifiers separate an input galaxy catalog into distinct groups defined by the specific algorithms, and output a file containing the integer group IDs for each galaxy, as well as the galaxy ID from the original catalog. One major usage of the Classifers is to divide the galaxy sample into tomographic bins. As in the other \texttt{RAIL.estimation} classes, Classifiers can take either the tabular catalog or a \texttt{qp.Ensemble} as input. 

The most straightforward Classifier algorithm is \texttt{Uniform\_binning}, which takes in a \texttt{qp.Ensemble} containing each galaxy's redshift PDF, and uses a single-valued point estimate redshift to assign the bin ID based on the bin edges provided. Similarly, \texttt{EqualCount} provides a method to separate the sample into bins of equal number counts, based on the provided redshift range and number of bins.
Objects that are not assigned into one of the bins are assigned with a special value which can be defined by the flag \texttt{no\_assign}.

The random forest classifier takes in a catalog containing the magnitudes of the galaxies.
This classifier, implemented in \code{rail\_sklearn}, uses the \code{sklearn.ensemble.RandomForestClassifier} to build a random forest using the training set magnitudes and colors, classifying the training galaxies into redshift bins set by the \code{zmin}, \code{zmax}, and \code{nzbins} configuration parameters.  \code{RandomForestClassifier} then finds the bin index from the trained redshift grid using the \code{predict} method of the \code{sklearn} classifier, and returns the tomographic bin or class index. {Note that the input of \code{RandomForestClassifier}, i.e., the photometric information used for training the model and for the classification, is flexible and can be a combination of magnitude and colors, magnitude-only, or color-only. The user decides the best set of input columns that suits their requirements. }




\begin{figure*}
  \centering
  \includegraphics[width=2.0\columnwidth]{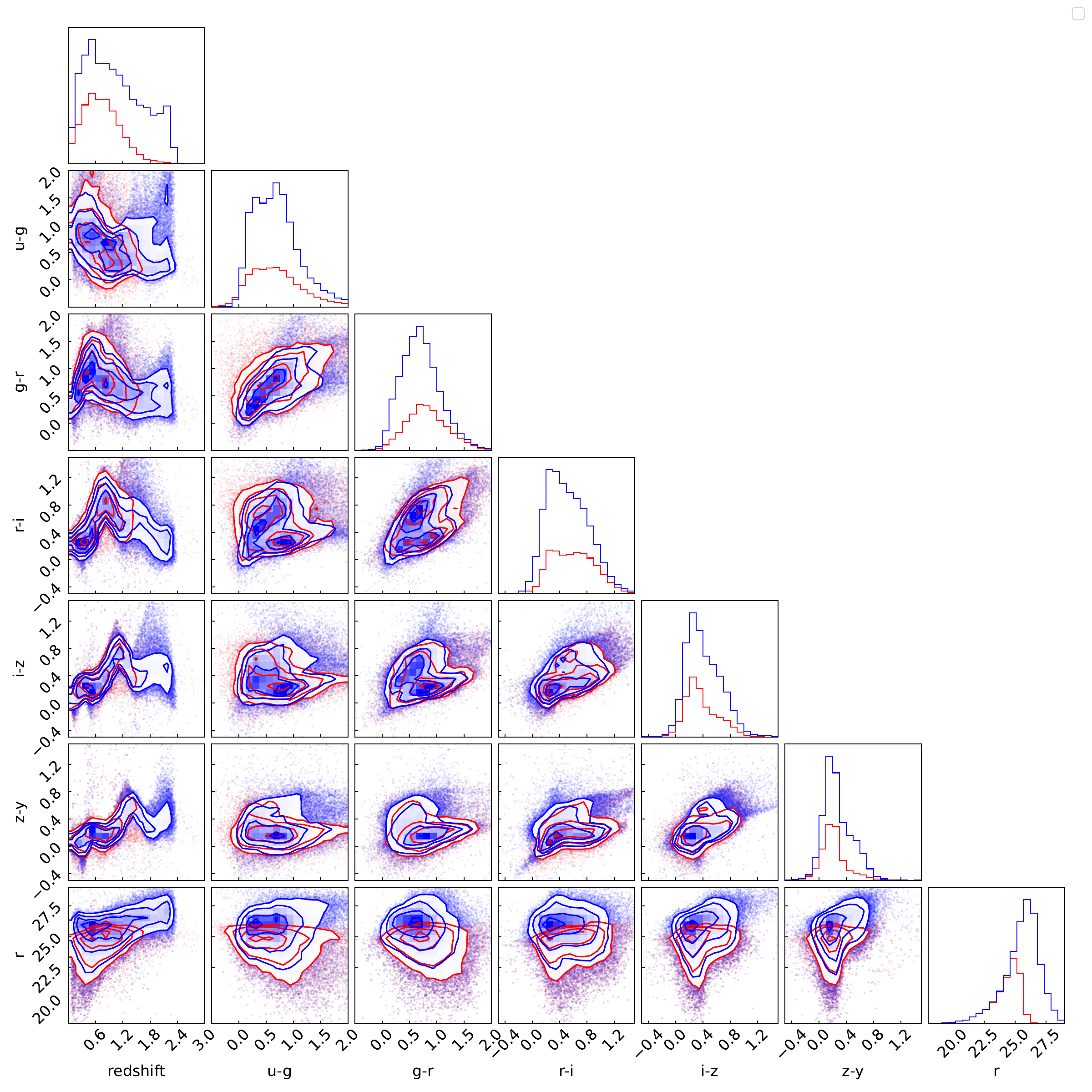}
  \caption{The color-redshift scatter of CosmoDC2 galaxies before (blue) and after (red) applying a series of degraders, which are described in Section~\ref{sec:examples:gs}. We can see that the population shown in red has a different distribution in color-redshift space compared to the population shown in blue.  }
  \label{fig:degrade} 
\end{figure*}

\section{Evaluation Modules}
\label{sec:eva}

RAIL provides evaluations of the performance of photo-$z$ methods through a library of metrics. {In Section~\ref{sec:dist-dist}, we introduce distribution-to-distribution metrics, which quantify the consistency between the photo-$z$ PDFs, $p(z)$, and the true redshift PDF, $p_{\rm true}(z)$ for the galaxy catalog. Section~\ref{sec:dist-point} presents distribution-to-point metrics that evaluate the performance of photo-$z$ PDFs against reference point estimates or true redshifts, $z_t$. Section~\ref{sec:point-point} shows point-to-point metrics that compare the point estimates of the galaxies with the truth. The base classes of these types are defined in \code{rail.evaluation}. In addition, in Section~\ref{sec:other-metrics}, we present a tomographic binning metric that calculates the overlapping fraction between two tomographic bins. 
}

\subsection{Distribution-to-Distribution Metrics}
\label{sec:dist-dist}

Distribution-to-distribution metrics compare the PDFs of the estimated galaxies with the `true PDF', which are the redshift conditional densities from which galaxies are drawn. These types of metrics can be more proper measures of the algorithm performance than those that take the true redshifts as point estimates. For instance, a galaxy with colors that can possibly be produced by a wide range of galaxies of different redshifts should be assigned a wide PDF on its estimate, even without noise. 

\begin{itemize}

\item Cram\'{e}r-von Mises (CvM): The Cram\'{e}r-von Mises (CvM) \cite{Cramer01011928} criterion defines the distance between the estimation and the truth by the cumulative density function (CDF),
\begin{equation}
{\rm CvM} = \int_{0}^{+\infty} [F_N(z) - F(z)]^2 \,p(z) dz,
\end{equation}
where $F(z)$ is the CDF of the observed redshift probability, and $F_N(z)$ is the CDF of the ``true distribution'' approximated by the empirical distribution, defined as 
\begin{equation}
    F_N (z) = \frac{N(z_i < z)}{N}.
\end{equation}
Here $N(z_i < z)$ is the number of observed redshift less than $z$, and $N$ is the total number of independent and identically distributed observations. The larger the CvM value the greater the likelihood that the estimate deviates. 

\item Kolmogorov-Smirnov (KS) test: Similar to the Cram\'{e}r-von Mises criterion, the Kolmogorov-Smirnov (KS) test defines the distance between the estimated probability and the true probability as the greatest difference between their CDFs,
\begin{equation}
{\rm KS} = {\rm sup}_z |F_N(z) - F(z)|.
\end{equation}
Since the CDF has a range of 0 to 1, the value of KS test can also range between 0 and 1, where 0 corresponds to perfect estimation.

\item Root Mean Square Error (RMSE): The Root Mean Square Error (RMSE) metric is computed as the RMSE integral between the estimated distributions $p(z)$ and the true distribution $p_{\rm true} (z)$
\begin{equation}
{\rm RMSE} = \sqrt{\int_{0}^{+\infty} (p(z) - p_{\rm true} (z))^2 \,dz}. 
\end{equation}
A high RMSE correspond to more statistically significant bias from the estimation. 

\item Kullback-Leibler Divergence (KL divergence): Kullback-Leibler Divergence (KL divergence) \cite{kl_divergence} defines the metric in terms of the relative entropy between the $p(z)$ and $p_{\rm true}(z)$, denoted $D_{\rm KL} (p || p_{\rm true})$
\begin{equation}
D_{\rm KL}(p||p_{\rm true}) = \int_{0}^{+\infty} p(z) \log\left(\frac{p(z)}{P_{\rm true}(z)}\right) \,dz.
\end{equation}
Note that the KL divergence does not commute, which means $D_{\rm KL}(p||p_{\rm true}) \neq D_{\rm KL}(p_{\rm true}||p)$. Similarly to the KS test, a higher KL divergence corresponds to a high degree of discrepancy between the estimation and the true distribution. 

\item Anderson-Darling (AD) test: The Anderson-Darling test is another metric based on the cumulative density functions of the estimated redshift and the true redshift correspondingly, similarly to the CvM. It is defined as 
\begin{equation}
{\rm AD} = N\sum_{i = 1}^{N} \int_{0}^{+\infty} \frac{(F_i(z) - F(z))^2}{F(z)(1-F(z))} \,p(z) dz,
\end{equation}
where $N$ is the number of galaxies and $F_i(z)$ is the empirical cumulative density function for the $i$-th galaxy. 

\end{itemize}

\subsection{Distribution-to-Point Metrics}
\label{sec:dist-point}
The Distribution-to-Point Metrics evaluate the performance of a photo-$z$ estimator on the resultant $p(z)$ against a reference point estimate or the truth.

\begin{itemize}
    \item Conditional density loss (\texttt{CDELoss}): we implement the method in \cite{2017arXiv170408095I} to compute the mean square error of the difference:
    \begin{equation}
        \label{eq:cdeloss}
        L = \mathbb{E} \left(\int (p^2(z|X) dz\right) - 2\mathbb{E} \left( p(z_t|X) \right), 
    \end{equation}
    where $z_t$ is true redshift, $X$ is the condition specific to the estimator, and the expectation is taken over the ensemble.  A better estimation should return a smaller CDE loss. The metric returns $L$ and the $p$-value of the difference. 
    \item Probability Integral Transformation (PIT): This is the cumulative distribution function of the photo-z PDF evaluated at the galaxy's true redshift for each galaxy in the catalog, i.e., 
    \begin{equation}
        {\rm PIT} = \int_{0}^{z_t} p(z) dz. 
    \end{equation}
    We provide a set of the PIT statistics such as quantiles, outlier rates, etc. One can also have access to the histogram of the PIT, which for a good estimation should be close to a uniform distribution. A tilted PIT histogram could indicate a biased PDF, or one that is under- or over-dispersed. Following the DC1 paper \citep{schmidt_evaluation_2020}, we provide the PIT-QQ (quantile-quantile) diagram, where the PIT distribution is directly compared to the ideal uniform distribution. A diagonal PIT-QQ diagram indicates a good estimation. 
    An example of the PIT-QQ plot is shown Section~\ref{sec:examples:gs}. 
    Furthermore, the metrics mentioned in Sec~\ref{sec:dist-dist} can also be used to assess how closely the PIT distribution is to the uniform $U(0,1)$.
    \item Brier Score: Proposed by Glenn Brier \citep{Brier1950}, The Brier score is a measure of the accuracy of probabilistic predictions. Given $N$ redshift PDFs, $p(z)$, characterised by $M$ redshift slices, the Brier score is defined as 
    \begin{equation}
        {\rm BS} = \frac{1}{N}\sum_{i=1}^{N}\sum_{j=1}^{M}(p_i(z_j)-\Delta(z_j-z_i^t))^2, 
    \end{equation}
    where $\Delta(z_j-z_i^t)$ is 1 when the true redshift of the $i$-th galaxy $z_i^t$ falls within the $j$-th slicing and 0 otherwise. $p_i(z_j)$ is the photo-$z$ probability within the $j$-th slice.  A lower Brier score indicates a more accurate distribution.
\end{itemize}

\subsection{Point-to-Point Metrics}
\label{sec:point-point}

The Point-to-Point metrics compare point estimates of the photo-$z$, $z_p$, computed by the PDF (this can be, e.g., the mean or the mode), to that of the truth or reference sample, $z_t$. The per-sample difference between the truth and photo-$z$ are first established by \texttt{PointStatsEz}, given by 
\begin{equation}
    \Delta_i = \frac{z_{p,i}-z_{t,i}}{1+z_{t,i}}.
\end{equation}
Several metrics are provided to characterise the distribution of $\Delta$:
\begin{itemize}
    \item \texttt{PointSigmaIQR} approximate the Gaussian standard deviation by the interquartile range (IQR) from 25-th percentile to 75-th percentile, i.e., the middle 50\% of the distribution of $\Delta$. The returned value the approximated Gaussian standard deviation $\sigma_{\rm IQR} = {\rm IQR}/1.349$;
    \item \texttt{PointBias} computes the median of \texttt{PointStatsEz};
    \item \texttt{PointOutlierRate} computes the fraction of the $\Delta$ distribution outside of $[0, {\rm max}(0.06, 3\sigma_{\rm iqr})]$. The upper limit is defined in the LSST Science Book \citep{lsstsciencecollaboration2009lsstsciencebookversion} and the lower bound of 0.06 is set to keep the fraction reasonable when $\sigma_{\rm IQR}$ is very small;
    \item \texttt{PointSigmaMAD} computes the standard deviation of the median absolute deviation (MAD), which is defined as 
    \begin{equation}
    {\rm MAD} = {\rm median}(|\Delta_i - {\rm median(\Delta)}|).
    \end{equation}
    The MAD is converted to the standard deviation via $\sigma_{\rm MAD} = 1.4862\,{\rm MAD}$.
\end{itemize}

\subsection{Other metrics}
\label{sec:other-metrics}

{The RAIL team is continuing to add more metrics into the codebase, especially those that are directly connected to specific science cases that utilize the photo-$z$ results. As an example, one important metric for clustering and weak lensing cosmology is the fraction of overlap between the two tomographic bins. A photo-$z$ algorithm that can better separate galaxies into tomographic bins improves the catalog's ability to trace the evolution of the large-scale structure.  We developed the \code{KDEBinOverlap} metric to compute the overlapping fractions between the $n(z)$ distributions for different tomographic bins, by approximating the $n(z)$ distributions using the kernel density estimation (KDE) on the true redshifts. The model produces an $N\times N$ matrix, where $N$ is the number of tomographic bins, with unity diagonal elements.}

\section{Examples and Tutorials}
\label{sec:examples and tutorials}

In this section, we showcase the key functionalities of \rail through a few examples and describe the tutorials available. The \rail tutorials are mostly located in the \code{rail\_hub}\footnote{\url{https://github.com/LSSTDESC/rail/tree/main/examples}}.  The `Golden Spike'\footnote{The Golden Spike is the ceremonial 17.6-karat gold final spike driven by Leland Stanford to join the rails of the first transcontinental railroad across the United States connecting the Central Pacific Railroad from Sacramento and the Union Pacific Railroad from Omaha on May 10, 1869, at Promontory Summit, Utah Territory \citep{wiki:Golden_spike}.} is a minimal `end-to-end' example of the \rail single object PDF and simple $N(z)$ estimation workflow, and is described in Section~\ref{sec:examples:gs}. Other tutorials are divided into Jupyter notebooks focused on degradation, estimation, and evaluation, as described in Section~\ref{sec:examples:others}. 

\subsection{The Golden Spike: an end-to-end demonstration of \rail}
\label{sec:examples:gs}

\begin{figure}
  \centering
  \includegraphics[width=1.0\columnwidth]{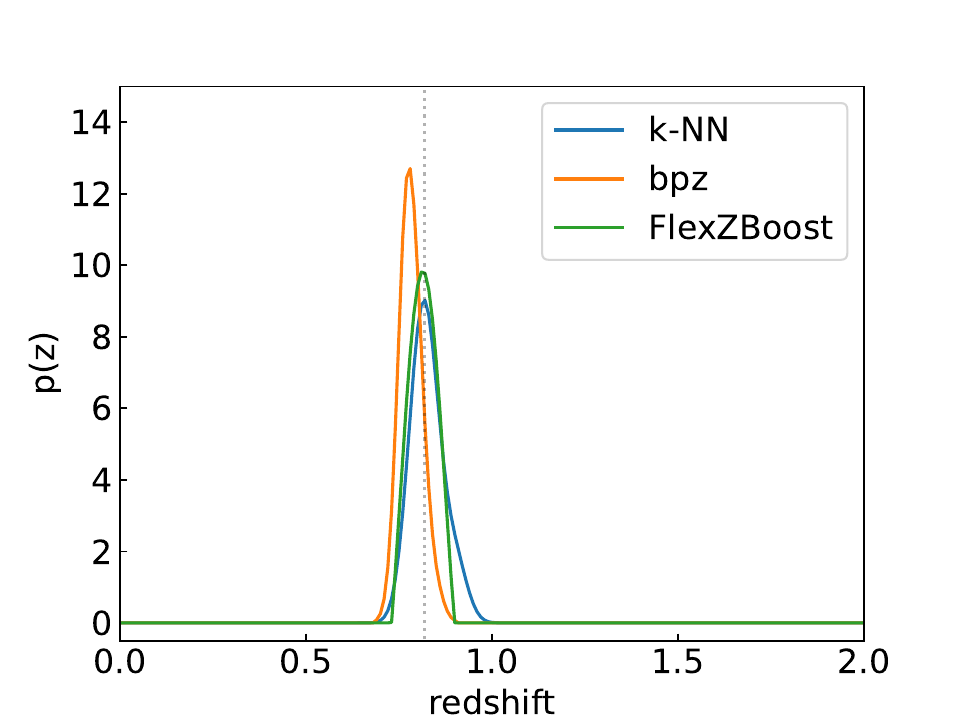}
  \caption{The probability density function of the redshift of a single galaxy from CosmoDC2, as estimated by three methods. The vertical line shows the true redshift.}
  \label{fig:estimation} 
\end{figure}

\begin{figure}
  \centering
  \includegraphics[width=1.0\columnwidth]{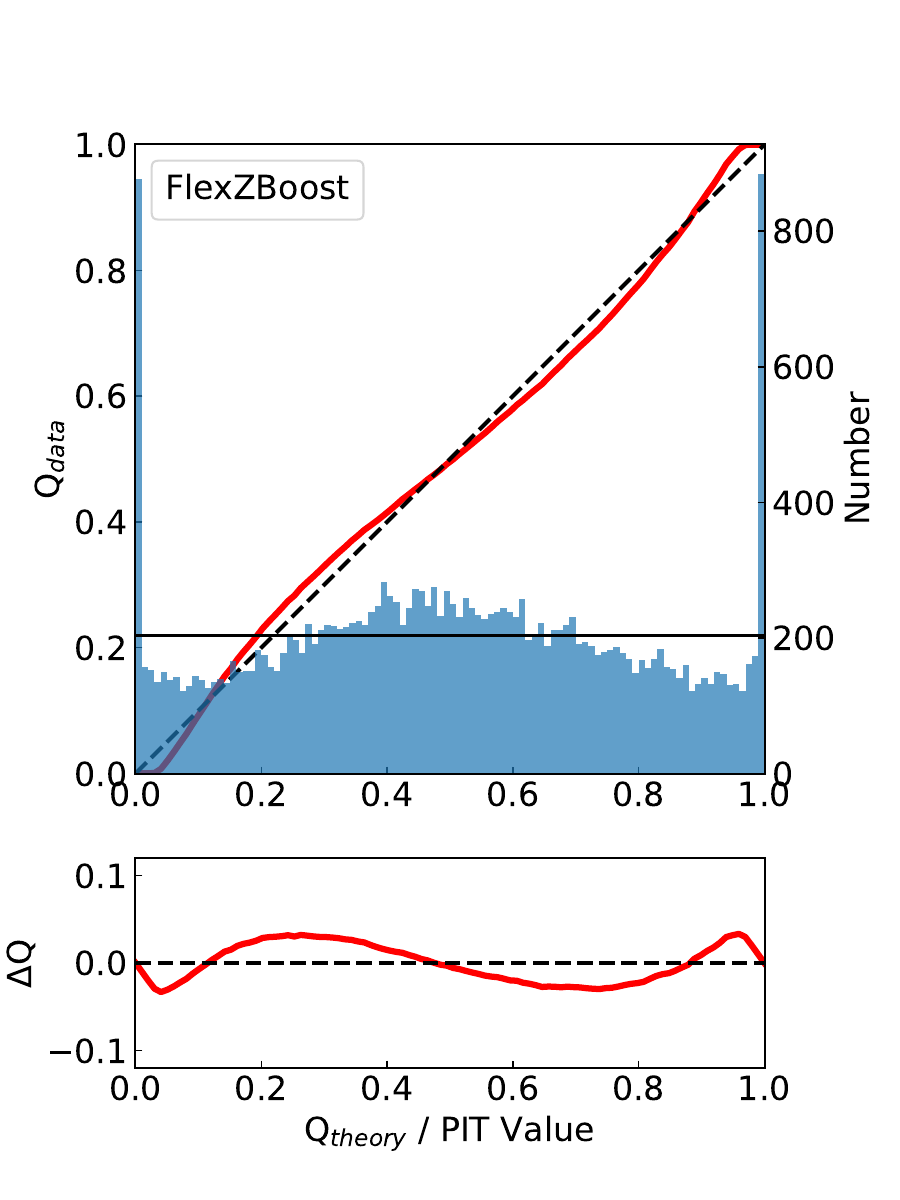}
  \caption{The Probability Transform Integral (PIT) metric evaluated for the \code{FlexZBoost} method as part of the Golden Spike, as compared with the ideal case (shown as the dashed lines). The metric shows that the estimator gives a relatively unbiased estimation of the redshifts and their uncertainties. Here $Q_{\rm data}$ denotes the quantile of the photo-$z$, while $Q_{\rm theory}$ is the quantile of the true redshift. $\Delta Q = Q_{\rm data} - Q_{\rm theory}$. {The spikes at 0 and 1 are due to outliers in the photo-$z$ estimates.}}
  \label{fig:evaluation} 
\end{figure}

The Golden Spike is a minimal end-to-end demonstration\footnote{\url{https://github.com/LSSTDESC/rail/tree/main/examples/goldenspike_examples}} of \rail's core functionality for estimating single object redshift PDFs and simple $N(z)$ ensemble estimates. The Golden Spike notebook has five main steps in the Golden Spike notebook: 
\begin{enumerate}
    \item Mock truth catalog generation: we train a flow model \citep{crenshaw2024b} to learn the mapping between the LSST photometry of a galaxy and its true redshift using a subset of the CosmoDC2 catalog \citep{Korytov_2019}. We then generate a set of mock galaxies with redshifts and LSST photometry. 
    \item Degradation of the truth catalog: we apply multiple degradation steps to the truth mock catalog in the following order: (a) the addition of analytical photometric noise to the photometry by the LSST error model, described in Section~\ref{sec:modules:lsst_error_model}; (b) an inverse redshift incompleteness selection with a pivot redshift of $z = 1.0$, using the \code{InvRedshiftIncompleteness} selector, (c) a 5\% OII-OIII line confusion degradation, via the \code{lineConfusion} degrader described in Sec~\ref{sec:modules:spectroscopic_degraders}, and (d) finally, an $i$-band magnitude cut at $i_{\rm mag} < 25$. In  Fig.~\ref{fig:degrade}, we show the color-redshift scattering of the mock catalog before and after the degradation, in blue and red contours respectively. 
    \item Training and estimation of photo-$z$: In the Golden Spike notebook, we split the degraded dataset into a training set and a testing set. We use the training set to train two photo-$z$ models: a $k$-nearest neighbor ($k$-NN, Section~\ref{sec:modules:knn}) and \texttt{FlexZBoost}, Section~\ref{sec:estimation:flexzboost}. We then apply \texttt{FlexZBoost}, $k$-NN, and \texttt{BPZ} (template-fitting code described in Section~\ref{sec:estimation:bpz}). In Fig.~\ref{fig:estimation}, we show the \pz PDFs of the three aforementioned methods for a random test galaxy. All three methods give comparable error estimates and are consistent with the true redshift. 
    \item Constructing the redshift distribution $n(z)$: The notebook creates a redshift distribution for the galaxy ensemble for each photo-$z$ method using the point estimate histogram method and the PDF stacking method, both of which are described in Section~\ref{sec:naive_methods}.
    \item Evaluating performance metrics: The Golden Spike notebook demonstrates \rail's ability to evaluate the performance metrics of the photo-$z$. We show the Probability Integral Transform function, a distribution-to-point metric described in Section~\ref{sec:dist-point}, in Fig.~\ref{fig:evaluation}. The PIT function (red line in the upper panel) is relatively close to the diagonal dashed line, which shows that \texttt{FlexZBoost} is providing photo-$z$ uncertainties that are consistent with the underlying uncertainties of the dataset. {Notice, however, that we do not set a `threshold' value for a metric to be `good enough'. We leave this assessment to follow-up studies that forward-model these impacts to cosmological analyses. }
\end{enumerate}

We note that Crafford et al. (\textit{in prep.}) use the framework of the Golden Spike and make a more in-depth study of the effect of many of degraders available in \rail, and their effects on photo-$z$ results via distribution-to-distribution metrics. 

\subsection{Other Examples}
\label{sec:examples:others}

We provide \textit{Jupyter} notebooks demonstrating individual functionalities for users who want to explore specific topics. These examples are organized into the following categories.

\subsubsection{Creation examples}

The creation examples demonstrate the use of \rail engines and degraders. They are organized in the following notebooks:
\begin{enumerate}
    \setcounter{enumi}{-1}
    \item \textbf{Quick Start in Creation} summarizes a series of degradations to a truth catalog. 
    \item \textbf{Photometric Realization} generates photometric realizations from different magnitude error models. 
    \item \textbf{Photometric Realization with Other Surveys} demonstrates the process of adding noise to the catalog with the photometric noise modules described in Section~\ref{sec:modules:lsst_error_model}, applied to LSST, Roman and Euclid bands.
    \item \textbf{Grid Selection for HSC} demonstrates the \code{GridSelector} described in Section~\ref{sec:modules:spectroscopic_degraders}.
    \item \textbf{Plotting Interface} shows the plotting interface with the DESC simulations, i.e., Skysim-5000, CosmoDC2 \citep{Korytov_2019}, and Roman-Rubin Simulation \citep{OpenUniverse}. 
    \item \textbf{True Posterior} demonstrates how to use \code{PZFlow}, as in Section~\ref{sec:pzflow_engine}, to calculate the true posteriors for a galaxy ensemble. 
    \item \textbf{Blending Degrader} demonstrates how to use the blending degrader described in Section~\ref{sec:blending}, and match the blended catalog to the truth catalog. 
    \item \textbf{DSPS SED} and 8. \textbf{FSPS SED} demonstrates some basic usage of the \code{rail\_dsps} and \code{rail\_fsps} library and described in Section~\ref{sec:dsps} and Section~\ref{sec:fsps}. 
    \setcounter{enumi}{8}
    \item \textbf{Spatial Variability} demonstrates the selection based on observing condition and described in Section~\ref{sec:observing_condition}.
    \item \textbf{Spectroscopic Selection for zCOSMOS} demonstrates the zCOSMOS spectroscopic selection from Section~\ref{sec:modules:spectroscopic_degraders}.
\end{enumerate}

\subsubsection{Estimation and summarization examples}

The estimation notebooks show how to use \rail's photo-$z$ estimation and summarization algorithms. 
\begin{enumerate}
    \setcounter{enumi}{-1}
    \item \textbf{Quick Start in Estimation} explains how to import a model for photo-$z$ estimation, and how to use that model to estimate $p(z)$.
    \item \textbf{FlexZBoost PDF Representation Comparison} explains the use of \code{FlexZBoost} (Section~\ref{sec:estimation:flexzboost}) and how to export the results in different statistical representations. 
    \item \textbf{BPZ lite} and 3. \textbf{BPZ lite with Custom SEDs} demonstrate the use of \texttt{BPZ}, described in Section~\ref{sec:estimation:bpz}.
    \setcounter{enumi}{3}
    \item \textbf{CMNN} demonstrates the use of the Color-Matched Nearest Neighbor, described in Section~\ref{sec:cmnn}.
    \item \textbf{DNF} demonstrates the use of the Directional Neighbor Fitting, described in Section~\ref{sec:dnf}.
    \item \textbf{GPz} demonstrates the use of \code{GPz}, as described in Section~\ref{sec:gpz}. 
    \item \textbf{NZDir} and 8. \textbf{NZDir pipeline} provide notebook and pipeline demo of the \code{NZDIR} estimator, described in Section~\ref{sec:nzdir}.
    \setcounter{enumi}{8}
    \item \textbf{PZFlow} demonstrates the use of \code{PZFlow}, described in Section~\ref{sec:pzflow}.
    \item \textbf{YAW} demonstrates the use of YetAnotherWizz \code{yaw}, described in Section~\ref{sec:yet_another_wizz}.
    \item \textbf{SomocluSom} and 12. \textbf{SomocluSOM Quality Control} demonstrate the use of the self-organizing maps implemented in \rail, described in Section~\ref{sec:som}, and the quality control measures implemented in \cite{Hildebrandt2021}.
    \setcounter{enumi}{12}
    \item \textbf{Sampled Summarizers} demonstrates the use of the summarizers, including the variational inference summarizer, naive methods in Section~\ref{sec:naive_methods}, and the \code{NZDir} summarizer in Section~\ref{sec:nzdir}.
\end{enumerate}

\subsubsection{Evaluation examples}

The evaluation examples demonstrate how to use the performance metrics implemented in \rail. They are organized in the following way:
\begin{enumerate}
    \setcounter{enumi}{-1}
    \item \textbf{Single Evaluator} demonstrates a single evaluator that evaluates all available metrics. 
    \item \textbf{Evaluation by Type} demonstrates separately the different metrics based on the types of statistics they evaluate, as listed in Section~\ref{sec:eva}.
\end{enumerate}

\subsubsection{Core examples}

The core examples are made to help \rail developers familiarize themselves with its data structure and developer tools. Because they are not user-facing, we do not elaborate here. However, we invite those interested in the detailed workings of \rail to review these notebooks. 

\subsection{Performance of \rail}
\label{sec:example:performance}

In this section, we briefly describe an initial study of the performance of \rail. We note that the goal of this work is not to systematically benchmark the performance of the \rail algorithms.
{There are ongoing efforts to compare the performance of all \rail algorithms with realistic incomplete spectroscopic training sample using mock LSST data via the $p(z)$ and $n(z)$ data challenge (The Dark Energy Science Collaboration \textit{et al. in prep.}). } 
{\rail is applied in Rubin Data Preview 1 in SITCOMTN-154 \citep{osti_2571480} and \cite{zhang2025photometricredshiftestimationrubin}. A systematic study of different metrics in the evaluation stage in \rail is conducted by Crafford \textit{et al. in prep.}.}

{We benchmark the memory usage of selective \rail algorithms, and summarize the maximum physical memory usage of each methods during training, estimation on  100,000 galaxies. We also record the model size of each method.  All methods use 6 bands magnitude and magnitude error as input. The results are shown in Table~\ref{tab:memmon}. We note that the memory usage of \texttt{kNN} and \texttt{LePhare} are relatively large during training. We defer the memory optimization of these methods to future development. }

\begin{table}[ht]
  \centering
  \begin{tabular}{lccc}
    \hline
    Algorithm & inform & estimate & model  \\
    \hline
    \texttt{BPz} & 0.177 & 0.401 & 0.0 \\
    \texttt{CMNN} & 0.121 & 0.167 & 8.4 \\
    \texttt{DNF} & 0.245 & 6.464 & 16.5 \\
    \texttt{FlexZBoost} & 0.601 & 0.804 & 45.8 \\
    \texttt{GPz} & 0.556 & 0.409 & 0.2 \\
    \texttt{kNN} & 4.936 & 2.089 & 6.2 \\
    \texttt{LePhare} & 11.842 & 1.078 & 413.7 \\
    \texttt{simpleNN} & 0.292 & 0.213 & 0.0 \\
    \texttt{TPz} & 0.462 & 0.858 & 87.1 \\
    \texttt{Train-z} & 0.118 & 0.314 & 0.0 \\
    \hline
  \end{tabular}
  \caption{Maximum physical memory (GB) and model size (MB) per algorithm and stage category.}
  \label{tab:memmon}
\end{table}

{
The level of the parallelization varies based on the type of stage. Most creation and degradation stages are not parallelized, since they are designed to process smaller datasets. The informers are generally also not MPI parallelized, as the photo-$z$ training set is usually not nearly as large as the LSST-scale catalog, and parallelization of training is usually specific to individual algorithms. 
As is pointed out in Section~\ref{sec:core:pipeline}, \rail estimation is parallelized to produce photo-$z$ at scale by reading the catalog in chunks. 
As an initial benchmarking, we summarize the estimation performance in Table~\ref{tab:speed} for \code{GPz}, \code{k-NN}, \code{FlexZBoost}, and \code{BPz}. Their speed can be extrapolated for multiple processes since MPI is deployed between the processes. This test is performed on the Rubin Science Platform hosted on the SLAC Shared Scientific Data Facility (S3DF). }

\begin{table}
    \centering
    \begin{tabular}{ll}
    \hline
        Algorithm & Evaluation speed [$(s\, \times{\rm CPU})^{-1}$]\\
    \hline
    \hline
        \code{BPz} & $2100$ \\
        \code{GPz} & $33000$ \\
        \code{k-NN} & $4000$ \\
        \code{FlexZBoost} & $1600$ \\
    \hline
    \end{tabular}
    \caption{The speed in terms of [galaxies$(s\, \times{\rm CPU})^{-1}$] for four photo-$z$ algorithms. The speed is rounded to the second digit. }
    \label{tab:speed}
\end{table}

\section{Summary}
\label{sec:end}

We present to the extragalactic astronomy community the Redshift Assessment Infrastructure Layers (\rail) software package, a comprehensive toolkit for end-to-end \pz pipelines. \rail was initiated and is developed by the \lsst-\desc, in collaboration with LINCC Frameworks. \rail is open-source, modular, and extensible, with intended usage throughout and beyond the Rubin ecosystem.
\rail's design welcomes contributions from the community, as models for generating mock photometry, algorithms for estimating redshifts and distributions thereof, and metrics of performance. This release represents a critical step toward ensuring that LSST photo-$z$ data products meet the stringent requirements of Rubin's cosmological and extragalactic science cases while also serving a broader community of researchers with varied scientific goals.

\rail enables studies that address key challenges identified in DESC's earlier photo-$z$ experiments, such as {discrepancies between algorithms}, inadequacies of traditional performance metrics, and the need for probabilistic approaches to model inherent redshift uncertainties \citep[e.g.,][Crawford et al. (\textit{in prep.})]{Moskowitz2024,hang2024,Merz2024}. It is built around three types of modules: creation, which provides tools for generating mock photometric catalogs with tunable imperfections and realistic complexities; estimation, which supports a unified API for implementing and comparing a diverse array of algorithms to compute per-galaxy and ensemble photo-$z$ PDFs; and evaluation, which offers a flexible suite of metrics, including principled mathematical measures and science-case-specific performance evaluations.

\rail provides many recent photo-$z$ algorithms, including machine learning, template fitting, hybrid, and image-based algorithms for per-galaxy photo-$z$ estimation.
\rail provides a common infrastructure for training models and estimating redshift PDFs which are parameterised by \texttt{qp}. The input/output is managed by \texttt{tables\_io}. 
\rail also provides algorithms to infer redshift distribution of an ensemble of galaxies, as well as algorithms to calibrate the redshift distributions, such as clustering redshift via \code{Yet\_Another\_Wizz}. Furthermore, \rail's modular structure facilitates extensibility, allowing users to integrate new methods, develop custom metrics, and adapt the framework to datasets beyond LSST.

The creation and evaluation modules in \rail provide valuable tools for photometric redshift research, especially towards comparing the performance of multiple photo-$z$ algorithms under a variety of circumstances. We expect that the \rail codebase will enable many algorithm comparison studies across different surveys and science cases. 

We demonstrate \rail's capabilities through practical examples, including the `Golden Spike' tutorial, which showcases an end-to-end workflow for generating mock catalogs, applying degradation, training photo-$z$ models, constructing redshift distributions, and evaluating their performance. This minimal demonstration highlights \rail's utility in stress-testing photo-$z$ methodologies and its ability to generate insights into systematic uncertainties.

Future development will focus on expanding the range of supported algorithms, incorporating feedback from early users, and addressing emerging challenges in photo-$z$ systematics. Efforts will include refining methods for incorporating emerging algorithms, handling corner cases, and exploring integration with Rubin commissioning pipelines.
For example, {a major planned addition} to \rail is the \code{SOMPZ} method \citep{2021MNRAS.505.4249M,Campos2024} adopted in the DES Y3 redshift calibration, which {uses a 2-tiered SOM to exploit deep field photometry and transfer the information to the wide field in a consistent, Bayesian approach. The resulting ensemble redshift distribution samples can also be combined with clustering redshift \citep[e.g.,][]{Giannini2024} to yield a joint constraint on $n(z)$.}
Concerning the integration with the rest of the LSST analysis pipeline, efforts are currently being made to propagate the uncertainties associated with the photometry estimates of the methods to cosmological constraints. {This is encompassed by the DESC package $\texttt{nz\_prior}$\footnote{\url{https://github.com/LSSTDESC/nz_prior}}, which maps samples of $n(z)$ from \rail to priors of chosen redshift uncertainty model, such as the `shift mean' model, and fully connects \rail with DESC likelihood codes such as $\texttt{firecrown}$. In a follow-up paper, {Ruiz-Zapatero \textit{et al.} (\textit{in prep.})} aims to report impact of different photometric uncertainty models on cosmology using $\texttt{nz\_prior}$.}
{Finally, There are ongoing efforts to validate and quantitatively compare the performance of different algorithms within RAIL under realistic conditions, including scenarios with incomplete and non-representative training data. These efforts are primarily carried out through LSST DESC internal data challenges, which explicitly address algorithm performance at both the per-object level $p(z)$ and the ensemble level $n(z)$. The outcomes of these challenges will be presented in a series of follow-up papers.}

By providing a flexible and extensible platform for photo-$z$ assessment, \rail aims to become a cornerstone of photometric redshift research, enabling precision cosmology on the new generation of photometric survey telescopes.

\subsection*{Data availability}

All \rail packages described in the paper are publicly available on GitHub. The example tutorial utilizes the CosmoDC2 dataset, which is also publicly available on \url{https://irsa.ipac.caltech.edu/Missions/cosmodc2.html}.


\subsection*{Contribution Statements}

%
%
J.L.~van den Busch: development of \texttt{yet\_another\_wizz} and its wrapper for integration in \rail, including optimizations for running computations on Rubin-like data sets. \\
E.~Charles: development of the core code, including the interfaces to other frameworks such as \code{ceci} and the Rubin data management software. Development of the data management model. Modularization of the code, implementation of various software pipelines. Supported code development within the \rail development team. \\
J.~Cohen-Tanugi: early software architecture development and documentation. Implementation of the interface to the \code{LePHARE} template photo-$z$ estimator.\\
A.~Crafford: testing the degradation, estimation and evaluation stages and providing feedback to the code development.\\
J.F.~Crenshaw: early development of the creation module, including normalizing flows, photometric error model, and spectroscopic degrades. \\
S.~Dagoret: adaptation of the \code{Delight} photometric redshift estimation code to the \rail interface.\\
J.~De-Santiago: parallelization of the estimators and summarizers.\\
J.~de Vicente: integration of Directional Neighbourhood Fitting (\code{DNF}) photo-$z$ on \rail framework.\\
Q.~Hang: development of the Observing Condition Degrader and classifiers, general software contributions, code and project administration, organization of telecon and discussion, writing and review of paper draft.\\
B.~Joachimi: provided guidance through mentorship of Q.~Hang and J.~Ruiz-Zapatero, along with detailed feedback on the manuscript.\\
S.~Joudaki: Development of the \code{Sphinx} documentation, and as photometric redshifts working group co-convener, created the \rail topical team and designed in-kind contributions towards the development of distinct aspects of \rail. \\
J.B.~Kalmbach: early development and conceptualization of the creation module, provided input on metrics. \\
A.~Kannawadi: Rubin Observatory builder; development of algorithms for galaxy colors measurement.\\
S.~Liang: development of the blending degrader \texttt{unrec\_bl\_model}. \\
O.~Lynn: general software development including improvements to scalability, code structure, documentation, and continuous integration. \\
A.I.~Malz: conceptualization, funding acquisition, investigation, methodology, project administration, software, supervision, validation, writing -- original draft, writing -- review \& editing.\\
R.~Mandelbaum: provided feedback on \rail development to members of the LINCC Frameworks team and through co-mentorship of A. Crafford, along with detailed feedback on the paper outline and text.\\
L. Medina-Varela: Memory profiling of \rail algorithms.\\
G.~Merz: code packaging, development of \code{rail\_deepdisc} and the description of the \code{DeepDISC} section of the paper.\\
I.~Moskowitz: development of \texttt{GridSelection} degrader. \\
D.~Oldag: code packaging, refactoring and optimization of core \rail components, metrics and estimation algorithms.\\
J.~Ruiz-Zapatero: development of the photometric uncertainity propagation pipeline, \texttt{nz\_prior}. \\
M.~Rahman: on behalf of Sidrat Research Inc, contributed to the development of underlying project infrastructure that facilitated these results, including \code{qp} and \code{tables\_io}.\\
M.M.~Rau provided the original implementation of NZDir summarizer, contributed to discussions, and co-convened of the PZ working group during parts of RAIL development.\\
S.J.~Schmidt: conceptualization and investigation, major software contributions, implemented many of the estimation stages for PZ algorithms, demo notebooks, code and project administration, writing and review of paper draft. \\
J.~Scora: on behalf of Sidrat Research Inc, contributed to the development of underlying project infrastructure that facilitated these results, including \code{qp} and \code{tables\_io}.\\
R.~Shirley: co-development of \code{rail\_lephare} code and writing overview of \code{rail\_lephare} for paper.\\ 
B.~St\"{o}lzner: co-development of spectroscopic degraders and \texttt{nz\_prior} \\
L.T.~San Cipriano: implemented the Directional Neighborhood Fitting (\code{DNF}) algorithm in the \rail system, enabling photometric redshift estimation using a nearest-neighbor approach that leverages directional information in magnitude space.\\
L.~Tortorelli: development of \texttt{rail\_fsps} and \texttt{rail\_dsps} in the creation module.\\
Z.~Yan: development of the creation module, including photometric error model and spectroscopic degrades; development of \texttt{rail\_som}. \\
T.~Zhang: write and review the paper draft, organize telecon and discussion, make major software contributions to various base classes and stages, and manage the codebase.\\ 

\section*{Acknowledgments}

This paper has undergone internal review in the LSST Dark Energy Science Collaboration. 
The authors would like to thank Seth Digel, Boris Leistedt, and Mike Jarvis for serving as the \desc publication review committee whose comments and suggestions that improved the quality of this manuscript.

A.I.M. and Z.Y. acknowledge support during this work from the Max Planck Society and the Alexander von Humboldt Foundation in the framework of the Max Planck-Humboldt Research Award endowed by the Federal Ministry of Education and Research. 

JLvdB is supported by an European Research Council Consolidator Grant (No. 770935).

L.T. acknowledges support from the Deutsche Forschungsgemeinschaft (DFG, German Research Foundation) under Germany's Excellence Strategy - EXC-2094 - 390783311.

S.L. is supported in part by the U.S. Department of Energy under grant number DE-1161130-116-SDDTA and under Contract No. DE-AC02-76SF00515 with the SLAC National Accelerator Laboratory. 

Q.H., J.R.Z. and B.J. are supported by STFC grant ST/W001721/1 and the UCL Cosmoparticle Initiative.

R.M. is supported in part by the Department of Energy grant DE-SC0010118.

The LINCC Frameworks team, including O.L., A.I.M., R.M., D.O., and T.Z., is supported by Schmidt Sciences. 

T.Z. thanks SLAC National Accelerator Laboratory for providing hospitality and an excellent research environment during the course of this study. 

G.M. is supported by LSST-DA through grant 2023-SFF-LFI-03-Liu,
NSF grant AST-2308174, and NASA grant 80NSSC24K0219.  G.M. thanks the LSST-DA Data Science Fellowship Program,
which is funded by LSST-DA, the Brinson Foundation, and
the Moore Foundation; his participation in the program has
benefited this work. G.M. also thanks Xin Liu for her support on this work as his PhD advisor.

Work at Argonne National Laboratory was supported by the U.S. Department of Energy, Office of High Energy Physics. Argonne, a U.S. Department of Energy Office of Science Laboratory, is operated by UChicago Argonne LLC under contract no. DE-AC02-06CH11357. MMR acknowledges the Laboratory Directed Research and Development (LDRD) funding from Argonne National Laboratory, provided by the Director, Office of Science, of the U.S. Department of Energy under Contract No. DE-AC02-06CH11357. Work at Argonne National Laboratory was also supported under the U.S. Department of Energy contract DE-AC02-06CH11357.

DESC acknowledges ongoing support from the IN2P3 (France), the STFC 
(United Kingdom), and the DOE and LSST Discovery Alliance (United States).  
DESC uses resources of the IN2P3 Computing Center 
(CC-IN2P3--Lyon/Villeurbanne - France) funded by the Centre National de la
Recherche Scientifique; the National Energy Research Scientific Computing
Center, a DOE Office of Science User Facility supported under Contract 
No.\ DE-AC02-05CH11231; STFC DiRAC HPC Facilities, funded by UK BEIS National 
E-infrastructure capital grants; and the UK particle physics grid, supported
by the GridPP Collaboration.  This work was performed in part under DOE 
Contract DE-AC02-76SF00515.

\bibliographystyle{mnras}
\bibliography{main}

\appendix

----------------------------------------------------------------------

\section{Utilities and Tools}
\label{sec:util_tools}


In this appendix, we describe supporting functionalities in \rail. These functionalities are divided into two main categories: utilities, which are classes and functions that facilitate easy access to catalog information and path-finding for local files across \rail, and tools, which are stages that provide some basic manipulation of the input catalog, such as reddening of fluxes and magnitudes.  

\subsection{Utilities}

\rail has two useful utilities. The \texttt{catalog\_utils} define several pre-set catalog-specific parameters, which can then be passed to methods as shared parameters. These parameters include photometry bands, magnitude limits, band name templates, reference bands, and the redshift column names. Pre-set catalogs include HSC, DC2 catalog, Rubin catalog, and the joint Roman-Rubin catalog (Troxel et al. In prep.). 

The other useful utility is \texttt{path\_utils}. Its functionality is useful when trying to retrieve the path to a particular file in \rail. By inputting the file name after \texttt{src/rail/}, the function returns the full path of the file in the system. 
This avoids issues where the paths can be different depending
on whether the code was installed from source. 

\subsection{Tools}

The \texttt{photometry\_tools} include a few useful \rail stages for photometry manipulation. Specifically, the \texttt{HyperbolicMagnitudes} allow the user to convert classical magnitudes and their respective errors to hyperbolic magnitudes \citep{Lupton_1999}, given a smoothing parameter that is estimated in the stage \texttt{HyperbolicSmoothing}. The \texttt{LSSTFluxToMagConverter} converts the LSST fluxes and their respective errors into magnitudes. 
The \texttt{Reddener} and \texttt{Dereddener} compute (de)reddened magnitudes given a dust map, the RA and Dec of the catalog objects, as well as the wavelength-dependent extinction factors for each band, $A_{\lambda}/E(B-V)$. The default dust map is the SFD map \citep{1998ApJ...500..525S,2011ApJ...737..103S} downloaded using the \code{Python} module \texttt{dustmaps}\footnote{\url{https://dustmaps.readthedocs.io/}}, but users can specify custom dust maps by specifying the paths to the maps.

The \code{table\_tools} functions provide several methods for tabular data manipulation. {The \code{ColumnMapper} re-maps the column names of the input catalog for, e.g., consistency with throughout the analysis}; the \code{RowSelector} sub-selects rows from a table by index; and the \code{TableConverter} converts tables from one format to another, e.g., from parquet to \code{Hdf5Table}.





\section{Dependencies}
\label{sec: dependencies}

\begin{table}
    \centering
    \begin{tabular}{ll}
    \hline
        Input file type & Data format\\
    \hline
    \hline
    Tabular data & \\
    
        FITS & Astropy table\\
        FITS & Numpy dictionary\\
        HDF5 & Astropy table\\
        HDF5 & Numpy dictionary\\
        HDF5 & Pyarrow table\\
        HDF5 & Pandas data frame\\
        Parquet & Pyarrow table\\
        Parquet & Pandas data frame\\
    \hline 
    PDF ensembles & \\
        \code{qp} &  \code{qp.Ensemble} \\
    \hline
    \end{tabular}
    \caption{File types handled by \code{tables\_io} and the \rail Data Handle.}
    \label{tab:tables_io}
\end{table}

\subsection{\tablesio}
\label{sec:tablesio}


\tablesio\footnote{\url{https://github.com/LSSTDESC/tables\_io}} was developed to abstract out the handling of catalog data from \rail and other \desc software, allowing such conversions to be performed automatically without the user having to manually preprocess data.
\tablesio enables \code{RailStage} objects to ingest data of a variety of formats, internally convert it to the format required by different wrapped engines and algorithms, and output it in a variety of formats, freeing the user from the responsibility to perform these conversions across \rail.


\subsection{\qp}
\label{sec:qp}

\qp\footnote{\url{https://github.com/LSSTDESC/qp}} is a \code{Python} package for handling univariate PDFs, originally introduced to optimize the storage parameterization for \lsst \pz PDFs \citep{malz_approximating_2018} but more recently refactored as a more flexible, scalable back-end for \pz PDFs provided in a variety of native formats and intended for different analyses.
All PDFs over redshift, be they catalogs of per-galaxy \pz PDFs or posterior samples of the redshift distribution of a sample of galaxies, are encapsulated by \qp.\code{Ensemble} objects, providing access to the same methods as any \code{scipy.stats.rv\_continuous} object \citep{scipy}, so the downstream consumers of the 1D PDFs need not hardcode a specific parameterization. 
In addition to all the \code{scipy.stats.rv\_continuous} parameterizations, \qp includes several additional parameterizations, such as spline interpolation, Gaussian mixture model, and grid histograms.
\qp also includes utilities for reducing PDFs to point estimates and for evaluating metrics of PDFs relative to other PDFs or scalar reference values, which are used as back-ends to the metrics of the \rail.\eva subpackage where applicable.

\section{Ecosystem}
\label{sec:eco}

\rail's functionality is spread across a constellation of GitHub repositories of the form \code{rail\_*}, but an installation\footnote{\url{https://rail-hub.readthedocs.io/en/latest/source/installation.html}} of \rail enables users to access the analogous functionality of all installed packages through a shared \code{Python} namespace.

All the standalone repositories depend on \code{rail-base}, which includes only the base classes, and are otherwise independent of one another.
This structure accommodates the desire of advanced users to install individual packages with the functionality that they need without waiting for additional, unrelated packages to install, and it protects new users from the risk of the entire installation failing in the case of even a single package breaking, which can happen if one of its dependencies introduces a breaking change.
In this paper, we refer to \rail from the user's perspective through the shared namespace, rather than referring to the set of standalone repositories.

\end{document}